\documentclass[10pt]{bmc_article}

\usepackage{graphicx}
\usepackage{subfigure}
\usepackage{amsmath}
\usepackage{amsfonts}
\usepackage{color} 
\usepackage{soul}
\usepackage{endnotes}

\usepackage{cite} 
\usepackage{url}  
\usepackage{ifthen}  
\usepackage{multicol}   
\usepackage[utf8]{inputenc} 
\newboolean{publ}

\newenvironment{bmcformat}{\baselineskip20pt\sloppy\setboolean{publ}{false}}{\baselineskip20pt\sloppy}

\begin{document}
\begin{bmcformat}


 \title{Identification of criticality in neuronal avalanches:\\ I. A theoretical investigation of the non-driven case}
 

\author{
	Timothy J Taylor$^1$%
         \email{Timothy J Taylor - 	tjt20@sussex.ac.uk}
        	\and
         	Caroline Hartley$^{2,3}$%
         	\email{Caroline Hartley - 	caroline.hartley@ucl.ac.uk}
       	\and
         	P\'eter L\ Simon$^{4}$%
         \email{P\'eter L\ Simon - 	simonp@cs.elte.hu}
       	\and
         	Istvan Z Kiss$^{5}$%
         \email{Istvan Z Kiss - 	I.Z.Kiss@sussex.ac.uk}
         \and
         	Luc Berthouze\correspondingauthor$^{1,2}$%
         \email{Luc Berthouze\correspondingauthor - L.Berthouze@sussex.ac.uk}
}


\address{%
    \iid(1) Centre for Computational Neuroscience and Robotics, University of Sussex, Falmer, Brighton BN1 9QH, UK\\
    \iid(2) Institute of Child Health, London, University College London, London WC1E 6BT, UK\\
    \iid(3) Centre for Mathematics and Physics in the Life Sciences and Experimental Biology, University College London, London, WC1E 6BT, UK\\ 
    \iid(4) Institute of Mathematics, E\"otv\"os Lor\'and University Budapest, Budapest, Hungary\\
    \iid(5) School of Mathematical and Physical Sciences, Department of Mathematics, University of Sussex, Falmer, Brighton BN1 9QH, UK
}%
 
\maketitle


\begin{abstract}
In this paper we study a simple model of a purely excitatory neural network that, by construction, operates at a critical point.  
This model allows us to consider various markers of criticality and illustrate how they should perform in a finite-size system. 
By calculating the exact distribution of avalanche sizes we are able to show that, over a limited range of avalanche sizes which we precisely identify, the distribution has scale free properties but is not a power law. 
This suggests that it would be inappropriate to dismiss a system as not being critical purely based on an inability to rigorously fit a power law distribution as has been recently advocated. 
In assessing whether a system, especially a finite-size one, is critical it is thus important to consider other possible markers. 
We illustrate one of these by showing the divergence of susceptibility as the critical point of the system is approached. 
Finally, we provide evidence that power laws may underlie other observables of the system, that may be more amenable to robust experimental assessment. 
\end{abstract}

\ifthenelse{\boolean{publ}}{\begin{multicols}{2}}{}



\section*{Introduction}

A number of \textit{in vitro} and \textit{in vivo} studies~\cite{bp1,bp2,petermann,hahn} have shown neuronal avalanches -- cascades of neuronal firing -- that may exhibit power law statistics in the relationship between avalanche size and occurrence. This has been used as \textit{prima facie} evidence that the brain may be operating near, or at, criticality~\cite{chialvo}. In turn, these results have generated considerable interest because a brain at or near criticality would have maximum dynamic range~\cite{kinouchi,shew,buckley} enabling it to optimally react and adapt to the dynamics of the surrounding environment~\cite{chialvo,lk01} whilst maintaining balanced neuronal activity~\cite{benayoun,magnasco,meisel12}. Neuropathological states (e.g., epileptic seizures) could then be conceptualised as a breakdown of, or deviation from, the critical state, see~\cite{milton}, for example. Furthermore, these findings have led to the notion that the brain may self-organise to a critical state~\cite{bak}, i.e., its dynamics would be driven towards the critical regime by some intrinsic mechanism and not be dependent on external tuning. In support of this view, Levina and colleagues~\cite{levina} showed analytically and numerically that activity-dependent depressive synapses could lead to parameter-independent criticality. 

The interpretation that neuronal activity is poised at a critical state appears to be mostly phenomenological whereby an analogy has been developed between the propagation of spikes in a neuronal network and models of percolation dynamics~\cite{essam} or branching processes~\cite{beggs,harris}. Remarkable qualitative similarities between the statistical properties of neuronal activity and the above models have given credence to this analogy, however, the question remains as to whether it is justified. Indeed, various key assumptions underlying percolation dynamics and branching processes are typically violated in the neuroscience domain. For example, full sampling, which is required in order to assess \textit{self-organised} criticality, is unattainable even in the most simple {\it in vitro} scenario and yet it has been shown that sub-sampling can have profound effects on the distribution of the resulting observations~\cite{priesemann}. On a related note, and more generally, the formal definition of a critical system as one operating at, or near, a second order (continuous) phase transition is problematic since the concept of phase transition applies to systems with infinite degrees of freedom~\cite{deco}. Many neuroscience authors address this by appealing to the concept of finite size scaling and many published reports implicitly assume that distributions are power law with truncation to account for the so-called finite size effect. Typically, such reports adopt an approach whereby (a) scale invariance is assessed through finite size scaling analysis, confirming that upon rescaling the event size, the curves collapse to a power law with truncation at system size (but see below regarding the definition of system size); (b) the parameters of statistical models are estimated, typically over a range of event size values that are rarely justified; and (c) the best model is determined by model selection, in which power law and exponentially truncated power law are compared to alternatives such as exponential, lognormal and gamma distributions, see~\cite{klaus} for a typical example. Whilst greater rigour in the statistical treatment of the assessment of the presence of power laws has been attained following Clauset and colleagues' influential paper~\cite{clauset}, what seems to be lacking is a rigorous treatment as to why a power law should be assumed to begin with. Although this question is particularly pertinent to the neurosciences, it should be noted that similar questions remain open in the field of percolation theory (e.g.,~\cite{ziff,chayes}), namely: (i) how does the critical transition behaviour emerge from the behaviour of large finite systems and what are the features of the transition? (ii) what is the location of the scaling window in which to determine the critical parameters? 

This paper specifically seeks to address the following questions: 
\begin{enumerate}
\item Assuming that the whole brain, or indeed a region of interest defined by where data can be obtained, is operating at criticality, can we reasonably expect power law statistics in neural data coming from a very small (possibly sub-sampled) subset of the system? If not, what would be the expected distribution? Sornette~\cite{sornette} states that the Gamma distribution is ``found in critical phenomena in the presence of a finite size effect or at a finite distance from the critical point". Jensen~\cite{jensen} claims that finite-size systems often show an exponential cut-off below the system size. However, we are not aware of any study in which the distribution of event sizes in a finite-size system set to operate at a critical regime has been investigated. 
\item In a finite-size system, is it reasonable/possible to perform a robust statistical assessment of power law statistics? Even the application of a rigorous model selection approach will lead to different results depending on the choice of the range of event sizes and the number of samples being considered~\cite{touboul}. The issue of range selection is of particular interest. Whilst the notion of system size is clear in models of criticality such as the Abelian sandpile where (i) there is full sampling, (ii) the number of sites is finite, and (iii) there is dissipation at the edges, system size is much less obvious where re-entrant connections exist, making it possible, in principle, for avalanches to be of infinite size. Here, the counting measure which leads to the definition of an avalanche is important. Counting the number of neurones involved in an avalanche will lead to a clearly defined system size, whereas counting the total number of activations -- the \textit{de facto} standard, e.g.,~\cite{beggsorg,levina,benayoun} -- will not. Furthermore, it should also be noted that the presence of re-entrant connections invalidates the standard theory of branching processes~\cite{harris}, and makes a rigorous determination of the branching parameter $\sigma$ problematic if not impossible, e.g., in the presence of avalanches merging. 
\item Are there other markers of criticality that may be more amenable to characterisation and that should be considered instead of, or in addition to, the statistics of event sizes? The need for such markers in neuroscience has been recognised (see~\cite{touboul} for example) and a number of studies have investigated long-range temporal correlations (power-law decay of the autocorrelation function) in amplitude fluctuations~\cite{linkenkaer2001} and in inter-burst intervals~\cite{hartley,segev}. However, a theoretical account of how those may relate to one another is lacking (although see the recent work in~\cite{poil}). Other markers of criticality (or markers of transitions) have been associated with critical physical systems, e.g., divergence of susceptibility and slowing of the recovery from perturbations near the critical point~\cite{sornette}, however, we are not aware of any theoretical or empirical study investigating them in a neuroscience context. 
\end{enumerate}

One way to address these questions more rigorously is to use simplified but therefore more tractable conceptual models (e.g., ~\cite{droste}). In this paper, we use a model of a purely excitatory neuronal system with simple stochastic neuronal dynamics which can be tuned to operate at, or near, a second order phase transition (specifically, a transcritical bifurcation). The simplicity of the model allows us to analytically calculate the exact distribution of avalanche sizes, which we confirm through simulations of the system's dynamics. We study our model at the critical point and compare our exact distribution to the explicit but approximate solution proposed by Kessler \cite{kessler} in an analogous problem of modelling disease dynamics. We confirm that Kessler's approximate solution converges to our exact result. Importantly, we show that, in the proposed finite-size system, this distribution is not a power law, thus highlighting the necessity of considering other markers of criticality. We thus analyse two potential markers of criticality: (i) the divergence of susceptibility that arises in the model as we approach the critical point, (ii) the slowing down of the recovery time from small disturbances as the system approaches the critical point. Finally, we speculate on a sufficient but not necessary condition under which our exact distribution could converge to a true power law in the limit of the system size.

\section*{The stochastic model}
We start from the stochastic model of Benayoun et al.~\cite{benayoun} which we simplify to the most trivial of models.  A 
fully connected network of $N$ neurones is considered with purely excitatory connections (as opposed to the excitatory and 
inhibitory networks considered in~\cite{benayoun}).    Within the network, neurones are considered to be either quiescent 
(Q) or active (A).  Taking a small time step $dt$ and allow $dt \rightarrow 0$ the transition probabilities between the 
two states are then given by:
\begin{align*}
	P\left(Q \rightarrow A, \text{ in time } dt\right) &= f\left(s_i(t)\right) dt \\
	P\left(A \rightarrow Q, \text{ in time } dt\right) &= \alpha dt
\end{align*}
where $s_{i}(t) = \sum_{j}{\frac{w_{ij}}{N}a_{j}(t)+h_i}$ is the input to the neurone.  Here $f$ is an activation function, 
$h_i$ is an optional external input, $w_{ij}$ is the connection strength from neurone $i$ to neurone $j$ and $a_j(t) = 1$ if 
neurone $j$ is active at time $t$ and zero otherwise. $\alpha$ is the de-activation rate and therefore controls the refractory period of the neurone.

To allow tractability, we further make the following simplifications:
\begin{enumerate}
\item We assume that all synaptic weightings are equal ($w_{ij} = w$).
\item We assume there is no external input. The driven case will be explored theoretically and empirically in a companion manuscript. 
\item We assume the linear identity activation function $f(x)=x$. Although it is more common to use sigmoid activation functions 
we note that the identity function can just be thought  of as a suitably scaled $tanh$ function over the desired range.
\end{enumerate}

As the network is fully connected we can write the following mean field equation for active neurones:
\begin{align*}
	\frac{dA}{dt} 	&= \frac{wA}{N}Q - \alpha A = \frac{wA}{N}(N-A) - \alpha A,
\end{align*}
where we have appealed to the fact that the system is closed and thus $A+Q=N$.  This ODE is analogous to the much studied~\cite{allen} 
susceptible $\rightarrow$ infectious $\rightarrow$ susceptible (SIS) model  used in mathematical epidemiology and we can appeal to some 
of the known results in studying its behaviour.  Primarily we can use simple stability analysis.  The non-zero steady state is given by 
$A^* = N(1-\alpha / w)$.  Setting $g(A) = \frac{dA}{dt}$, this equilibrium point is stable if $g'(A^*) < 0$.  Thus
\begin{align*}
	g'(A) = (w-\alpha)-2w\frac{A}{N} \Rightarrow g'(A^*) = (w-\alpha)-2w\frac{N(1-\alpha / w)}{N} = \alpha - w.
\end{align*}

Borrowing from epidemiology we define the threshold $R_{0} = \frac{w}{\alpha}$.  If $R_{0}>1$ we see that  
$g'(A^*) = \alpha - w < 0$ and the non-zero steady state is stable.  Figure~\ref{fig1} illustrates the differing behaviour of the solution
to the above ODE for $R_0 < 1$ (sub-critical), $R_0 = 1$ (critical) and $R_0 > 1$ (super-critical). 
\begin{figure}[!ht]
    \begin{center}
        \includegraphics[width=0.7\textwidth]{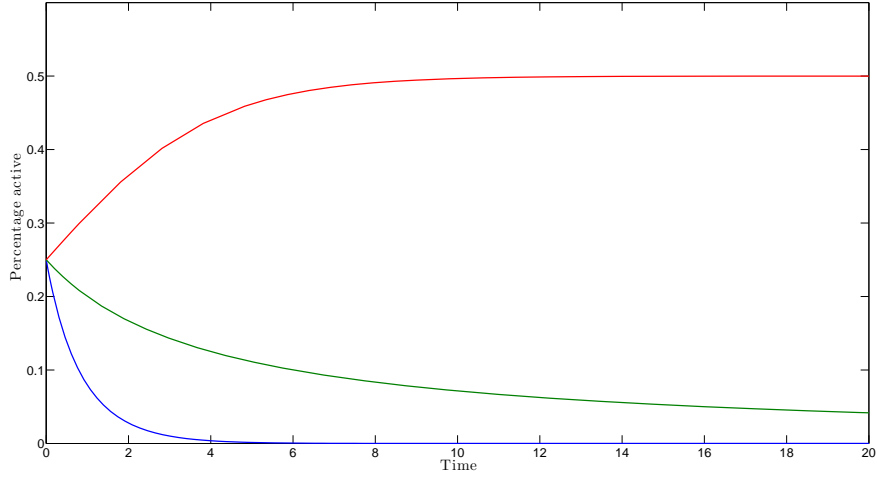}
    \end{center}
    \caption{
    {\bf Activity in the different regimes.} Plot of the solution to the ODE for $N = 800$ and three different regimes where $R_0$ was 
    set to $0.5$ (blue), $1.0$ (green) and $2.0$ (red). Initially we activated 25\% of the network.}
    \label{fig1}
\end{figure}
\subsection*{Firing neurones and avalanches}
Instead of focussing on the average activity level across the network we will instead look at the size distribution of firing neurones following 
one firing event. To do this we begin with a fully quiescent network and initially activate just one neurone.  We then record the total number 
of neurones that fire (the number of quiescent to active transitions) until the network returns to the fully quiescent state.  We use this process of sequential activation as our definition of an 
avalanche and our main interest is the distribution of the avalanche sizes.  Unfortunately, the simple ODE approach will not provide us 
with this distribution. To calculate this distribution, we use the semi-analytic approach described in the following section.
\subsection*{Tree approach to the avalanche distribution}
\label{treeapproach}
We begin by defining $q_i$ as the probability the next transition is a recovery (from A to Q) given $i$ active neurones ($i>0$).  The 
probability the next transition is an activation is then $1-q_i$ and we have:
\begin{align*}
	q_i 		&= \frac{\alpha N}{w(N-i)+\alpha N} = \frac{N}{R_0(N-i) + N},\\
	1-q_i	&= \frac{w(N-i)}{w(N-i)+\alpha N} = \frac{R_0(N-i)}{R_0(N-i) + N}.
\end{align*}
In order to calculate the avalanche size distribution we adopt a recursive approach.  We begin by considering the process unfolding in a 
tree like manner with $1$ initially active neurone.  The tree can be divided into levels based on how the process is unfolding.  Each level 
is itself subdivided into two containing the possible number of active neurones  before and after a transition has occurred.  Level $k$ 
contains firstly the number of active neurones after $2k$ transitions and secondly all possible active neurones after $2k+1$ transitions.  
The recursive tree approach relates the probability of transition between and within these levels to the final avalanche size.  
Figure~\ref{fig2} illustrates the first $3$ levels of this process.
\begin{figure}[!ht]
		\begin{center}
				\includegraphics[width=0.7\textwidth]{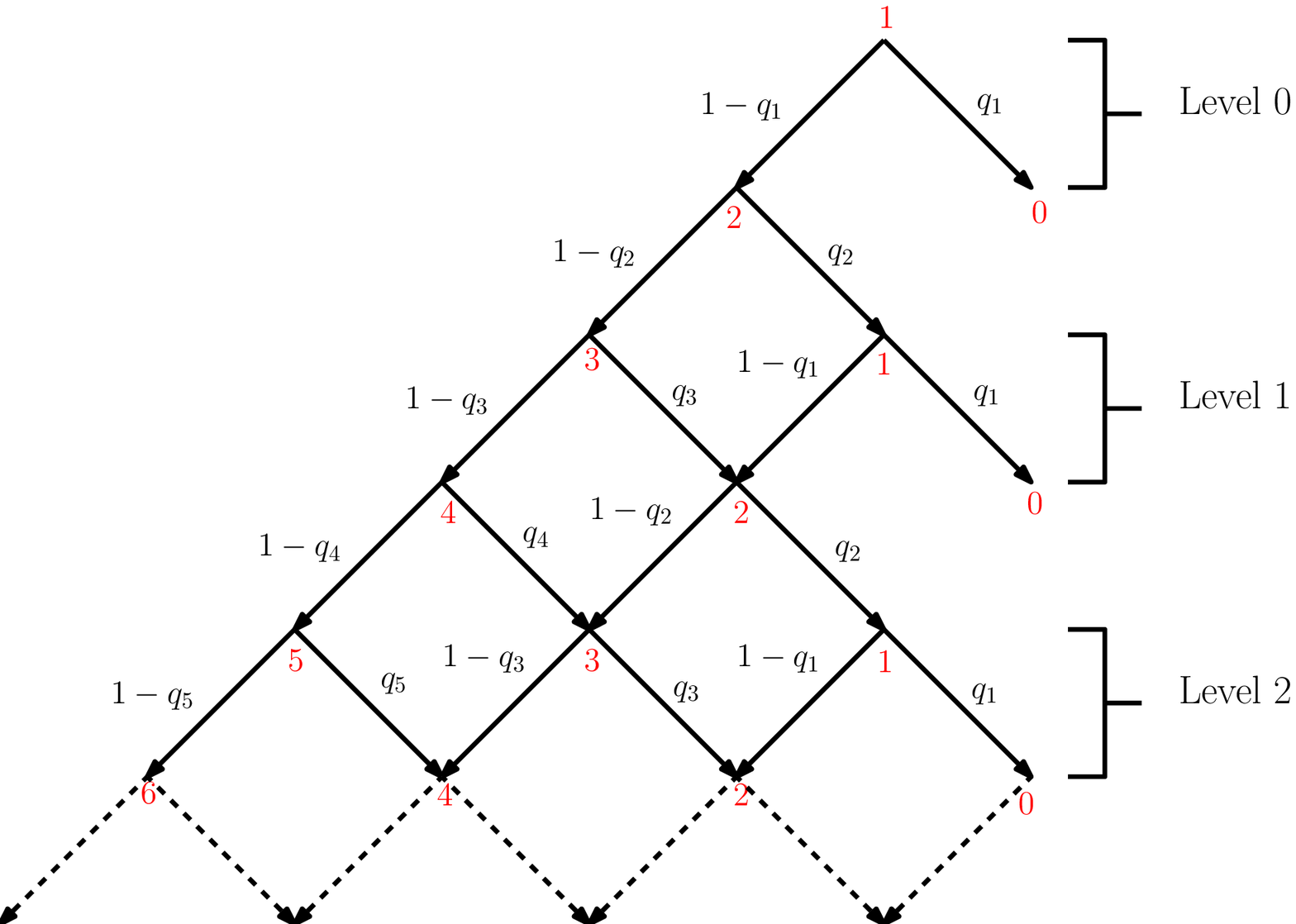}
		\end{center}
		\caption{
			{\bf First three levels of the probability tree.} Red numbers are the number of active neurones, black values
			are the probability of transitions between levels and sub levels.
			}
\label{fig2}
\end{figure}
To continue we define $p_k^i$ as the probability of having $i$ active neurones on level $k$ with $i = 0,1,2, \ldots ,N$ and $k \in \mathbb{N}_0$.  
On the first level ($k = 0$) we immediately see that $p_0^1 = 1$, $p_0^2 = 1-q_1$ and $p_0^0 = q_1$.  To proceed we will consider the 
probability of having an odd number of active neurones on an arbitrary level.  First we note the following within level recurrence
\begin{align*}
	p_k^{2j} 			&= p_k^{2j-1}\left(1-q_{2j-1}\right) + p_k^{2j+1}q_{2j+1} \quad (j \geq 1).
\end{align*}
\\ \\
We can now use these in our calculation of $p_{k+1}^{2j+1}$ and obtain the following between level recurrence
\begin{align*}
	p_{k+1}^{2j+1}	= &p_k^{2j}\left(1-q_{2j}\right) + p_k^{2j+2}q_{2j+2} \\
					= &p_k^{2j-1}\left(1-q_{2j-1}\right)\left(1-q_{2j}\right) + p_k^{2j+1}q_{2j+1}\left(1-q_{2j}\right) +p_k^{2j+1}\left(1-q_{2j+1}\right)q_{2j+2} + p_k^{2j+3}q_{2j+3}q_{2j+2} \\
					= &p_k^{2j-1}\left(1-q_{2j-1}\right)\left(1-q_{2j-1}\right) + p_k^{2j+1}\left(q_{2j+1}\left(1-q_{2j}\right) +\left(1-q_{2j+1}\right)q_{2j+2}\right) + p_k^{2j+3}q_{2j+3}q_{2j+2}.
\end{align*}

This newly derived recurrence relation provides a consistent set of equations between levels in terms of odd numbers of active neurones.
This algebraic manipulation was necessary to obtain a self-consistent system of equations.  Letting $r_{j}(k) = p_k^{2j+1}$ we have:
\begin{align*}
	r_j(k+1) = &r_{j-1}(k)\left(1-q_{2j-1}\right)\left(1-q_{2j}\right) + r_{j}(k)\left(q_{2j+1}\left(1-q_{2j}\right) +\left(1-q_{2j+1}\right)q_{2j+2}\right) + r_{j+1}(k)q_{2j+3}q_{2j+2},
\end{align*}
where we must make suitable modifications for systems with only an individual neurone active and also near
and at full activation.  For a single active neurone the equation is simply given as
\begin{align*}
r_{0}(k+1) = &r_{0}(k)(1-q_{1})q_2 + r_{1}(k)q_{3}q_{2}.
\end{align*}
For full (or near full) activation the equations depend on the system size, $N$.  Defining $\tilde{N} = (N-2)/2$ if $N$ even
and $\tilde{N} = (N-1)/2$ for $N$ odd we then obtain
\begin{align*} 
r_{\tilde{N}}(k+1)=\begin{cases}
    r_{\tilde{N}-1}(k)\left(1-q_{2\tilde{N}-1}\right)\left(1-q_{2\tilde{N}}\right) +
    r_{\tilde{N}}(k)\left(q_{2\tilde{N}+1}\left(1-q_{2\tilde{N}}\right) + \left(1-q_{2\tilde{N}+1}\right)q_{2\tilde{N}+2}\right), 
    & \text{if $N$ even},\\
    r_{\tilde{N}-1}(k)\left(1-q_{2\tilde{N}-1}\right)\left(1-q_{2\tilde{N}}\right) +
    r_{\tilde{N}}(k)q_{2\tilde{N}+1}\left(1-q_{2\tilde{N}}\right) , & \text{otherwise}.
  \end{cases}
\end{align*}
\\ \\
We now define:
\begin{align*}
\mathbf{r}(k)  &= \left(\begin{array}{c}
				r_0(k) \\
				. \\
				. \\
				. \\
				. \\
				. \\
				r_{\tilde{N}}(k)
			\end{array}\right).
\end{align*}
We can now write $\mathbf{r}(k+1) = \mathbf{A}\cdot\mathbf{r}(k)$ where, for N even, matrix $\mathbf{A}$ 
is given by the following tridiagonal matrix:
\begin{align*}
	\mathbf{A}  &= \left(\begin{array}{ccccccccc}
				\left(1-q_1\right)q_2 & q_3q_2 & & & & & \\
				& \ddots &\ddots & & & & \\
				& & \ddots & \ddots & & & \\
				& & b_{j} & c_{j} & d_{j} & & \\
				& & & \ddots &  & & \\
				& & & & \ddots &  &  \\
				& & & & & \left(1-q_{2\tilde{N}-1}\right)\left(1-q_{2\tilde{N}}\right) & q_{2\tilde{N}+1}\left(1-q_{2\tilde{N}}\right) + q_{2\tilde{N}+2}\left(1-q_{2\tilde{N}+1}\right)
			\end{array}\right)
\end{align*}
where 
\begin{align*}
	b_{j} &= \left(1-q_{2j-1}\right)\left(1-q_{2j}\right)\\
	c_{j} &= q_{2j+1}\left(1-q_{2j}\right)+q_{2j+2}\left(1-q_{2j+1}\right) \\
	d_{j} & = q_{2j+3}q_{2j+2} \qquad\qquad\qquad\qquad\qquad\qquad \text{and $j = 1:(\tilde{N}-1)$.}
\end{align*}

On the $k^{th}$ level of the tree, the probability of only $1$ neurone being active is given by $p_{k}^{1} = r_{0}(k)$.  
We can then calculate the probability of zero active neurones after $k$ firings as $q_{1}r_0(k)$; this is thus
the probability, $P(k+1)$, of having avalanches of size $k+1$ since initially one neurone is active.  
To calculate the distribution we implemented the recursive method of calculation in the MATLAB\textsuperscript{\textregistered} environment.
\subsection*{Simulations of neuronal avalanches}
In order to check the validity of our method, we performed simulations of the firing neurones using the Gillespie algorithm~\cite{gillespie}.  
Due to the network being fully connected the algorithm is relatively straightforward:
\begin{itemize}
	\item 	As earlier let $A$ be the number of active neurones in the network ($Q$ the number of quiescent).  Given that an 
			individual neurone becomes quiescent at rate $\alpha$ then the total rate of (Active $\rightarrow$ Quiescent) 
			transitions is given by $r_{aq} = A\alpha$. Similarly the total rate of (Quiescent $\rightarrow$ Active) transitions is given by $r_{qa} = f(s_i)Q = f(s_i)(N-A)$.
	\item	Let $r = r_{aq} + r_{qa}$ and generate a timestep $dt$ from an exponential distribution of rate $r$.
	\item	Generate a random number n between $0$ and $1$. If $n < \frac{r_{aq}}{r}$ an
			active neurone turns quiescent, otherwise a quiescent neurone is activated (fires).  This event is said
			to occur at time $t+dt$ and the network is updated accordingly.  
\end{itemize}
\subsection*{Exact solution compared to simulation}
Values of the threshold, $R_{0}$, were chosen less than, equal to and finally above $1$. We will refer to these regimes as subcritical, 
critical and supercritical respectively.  Figure~\ref{fig3} illustrates the, as expected, good agreement between the simulations and the 
exact result for the three different regimes of $R_{0}$.
\begin{figure}[!ht]
		\begin{center}
				\includegraphics[width=0.7\textwidth]{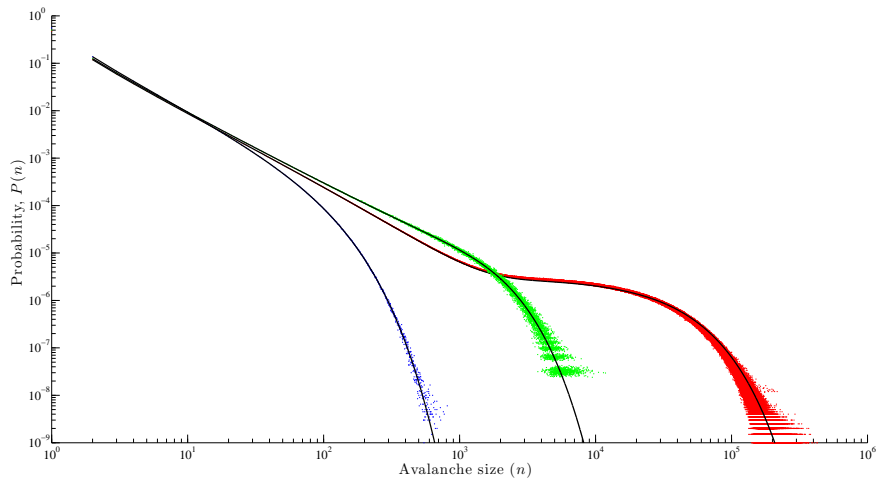}
		\end{center}
		\caption{
			{\bf Avalanche distributions.} Results from the simulations of the avalanche distributions for the subcritical ($R_{0} < 1$, blue), critical ($R_{0} = 1$, green) and supercritical ($R_{0} > 1$, red) regimes for a network of size $N=800$.  For each regime $2,000,000,000$ avalanches were simulated.  The corresponding exact solutions are shown in black.
			}
\label{fig3}
\end{figure}
\subsection*{A closed solution}
In~\cite{kessler}, Kessler proposed a closed solution to the analogous susceptible-infected-susceptible (SIS) problem 
where he was interested in the number of infections (including reinfections) occurring over the course of an epidemic.  
For small avalanche sizes where the number of infectives is negligible compared to the network size the transition 
probabilities can be approximated as
\begin{align*}
	q_i 		&= \frac{N}{R_0(N-i) + N} \approx \frac{1}{R_0 + 1},\\
	1-q_i	&= \frac{R_0(N-i)}{R_0(N-i) + N} \approx \frac{R_0}{R_0 + 1}.
\end{align*}
In the critical regime $R_0 = 1$, the problem reduces to calculating the distribution of first passage times of a 
random walk with equal transition probabilities.  Thus for avalanche sizes in the range, $1 \ll n \ll \sqrt{N}$, Kessler
\cite{kessler} gave the following distribution based on Stirling's approximation
\begin{align}
P(n) &= \frac{1}{2^{2n-1}}\left[\binom{2n-2}{n-1} - \binom{2n-2}{n}\right] \approx \frac{1}{\sqrt{4 \pi n^3}}.
\label{eq1}
\end{align}
We note however that the range over which the distribution can be shown to be a power law is rather limited and for small networks will not hold.
Using the theory of random walks and a Fokker-Planck approximation Kessler also derived the following closed solution to the probability distribution of infections in the critical regime ($R_{0}=1$) for larger sizes:
\begin{align}
P(n) &= \frac{1}{\sqrt{4\pi N^3}}\exp(n/2N)\sinh^{-\frac{3}{2}}\left(n/N\right) \quad (n \gg 1)
\label{eqkessler}
\end{align}
Figure~\ref{fig4} plots this approximation against our exact solution for a network of size $N = 800$.  To more formally assess the convergence of the approximate solution to that of our exact solution we considered the probabilities of avalanches from size $N/10$ to $20N$ 
and measured the difference between the distributions using two different metrics.  Letting $P_e(n)$ be the exact probability of an avalanche of size $n$ and $P_k(n)$ be the Kessler approximation to this we first considered the standard mean-square error given by
\begin{align*}
Error(N) = \frac{1}{R}\sum_{n=N/10}^{20N}\left(P_e(n) - P_k(n)\right)^{2} \quad \text{where }R = 20N - N/10 + 1.
\end{align*}
Secondly we considered a more stringent measure of the error by looking at the supremum of difference between the same range of avalanches
\begin{align*}
Error(N) = \sup_{n}|P_e(n)-P_k(n)|.
\end{align*}
Figure~\ref{fig5} illustrates the two errors for increasing network size and both show how the proposed closed solution is indeed converging to that of the exact.  
\begin{figure}[!ht]
		\begin{center}
				\includegraphics[width=0.7\textwidth]{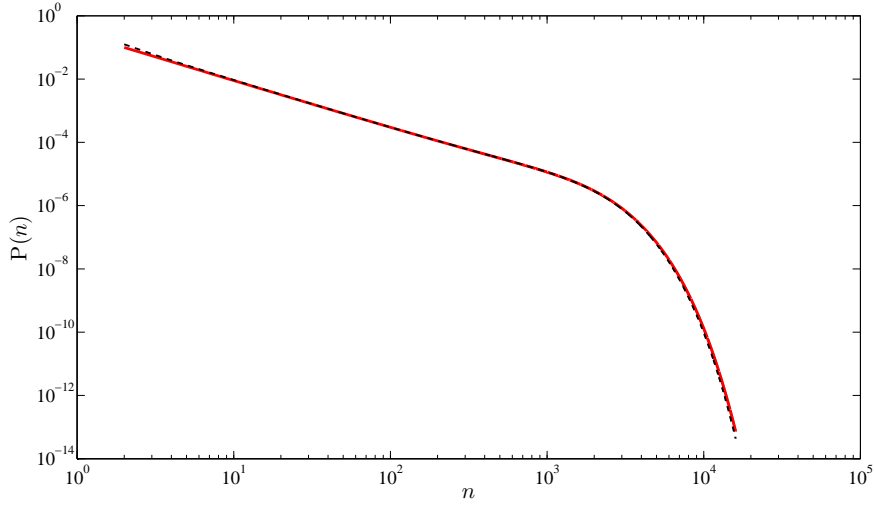}
		\end{center}
		\caption{
			{\bf Closed solution versus exact.} Plot of the closed solution (red solid line) versus the exact solution (black dashed line) for a network of size $N=800$ operating in the critical regime.
			}
\label{fig4}
\end{figure}
\begin{figure}[!ht]
    \begin{center}
        \subfigure[Mean square error]{
            \label{fig5a}
            \includegraphics[width=0.45\textwidth]{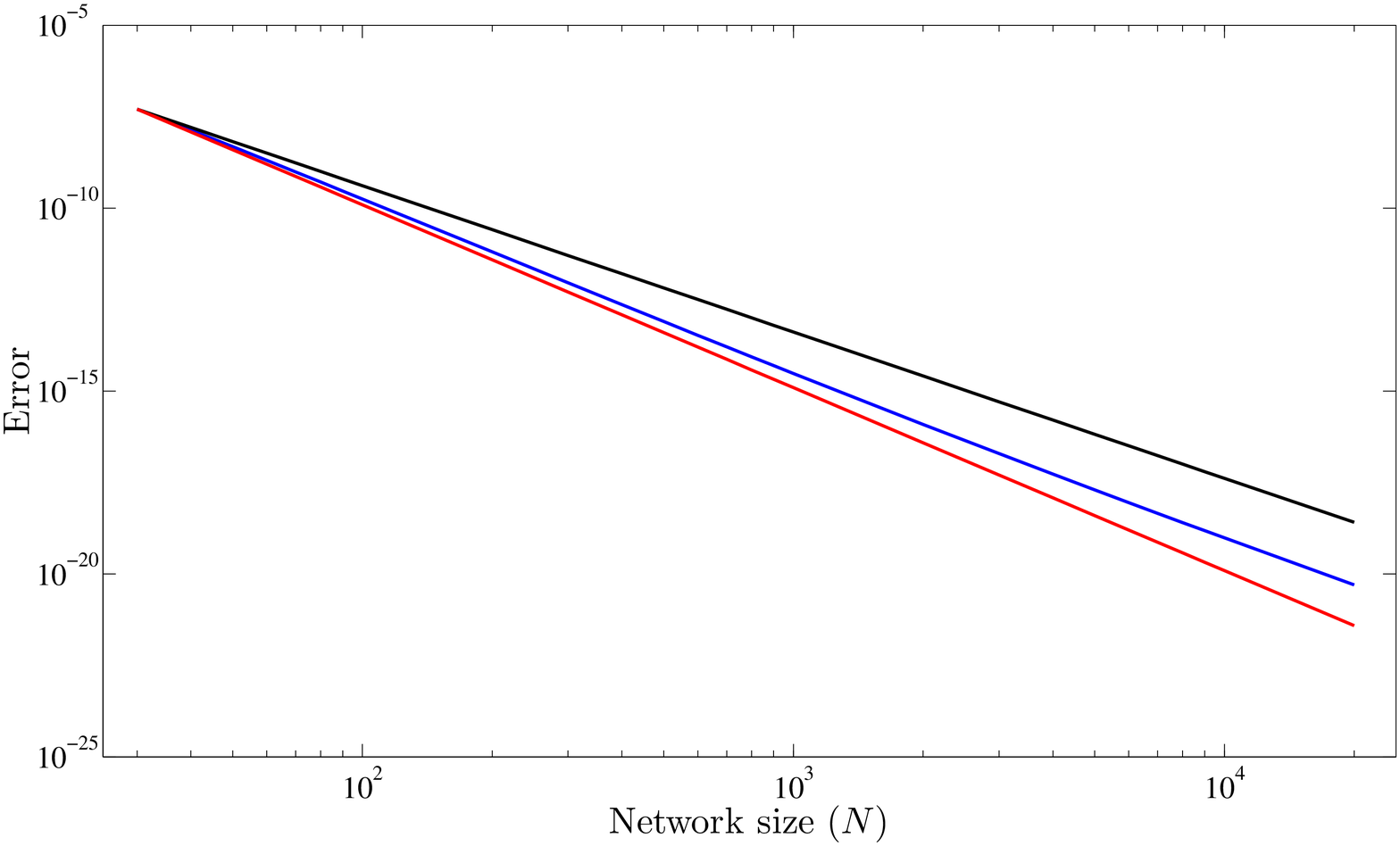}
        }
        \subfigure[Supremum error]{
           \label{fig5b}
           \includegraphics[width=0.45\textwidth]{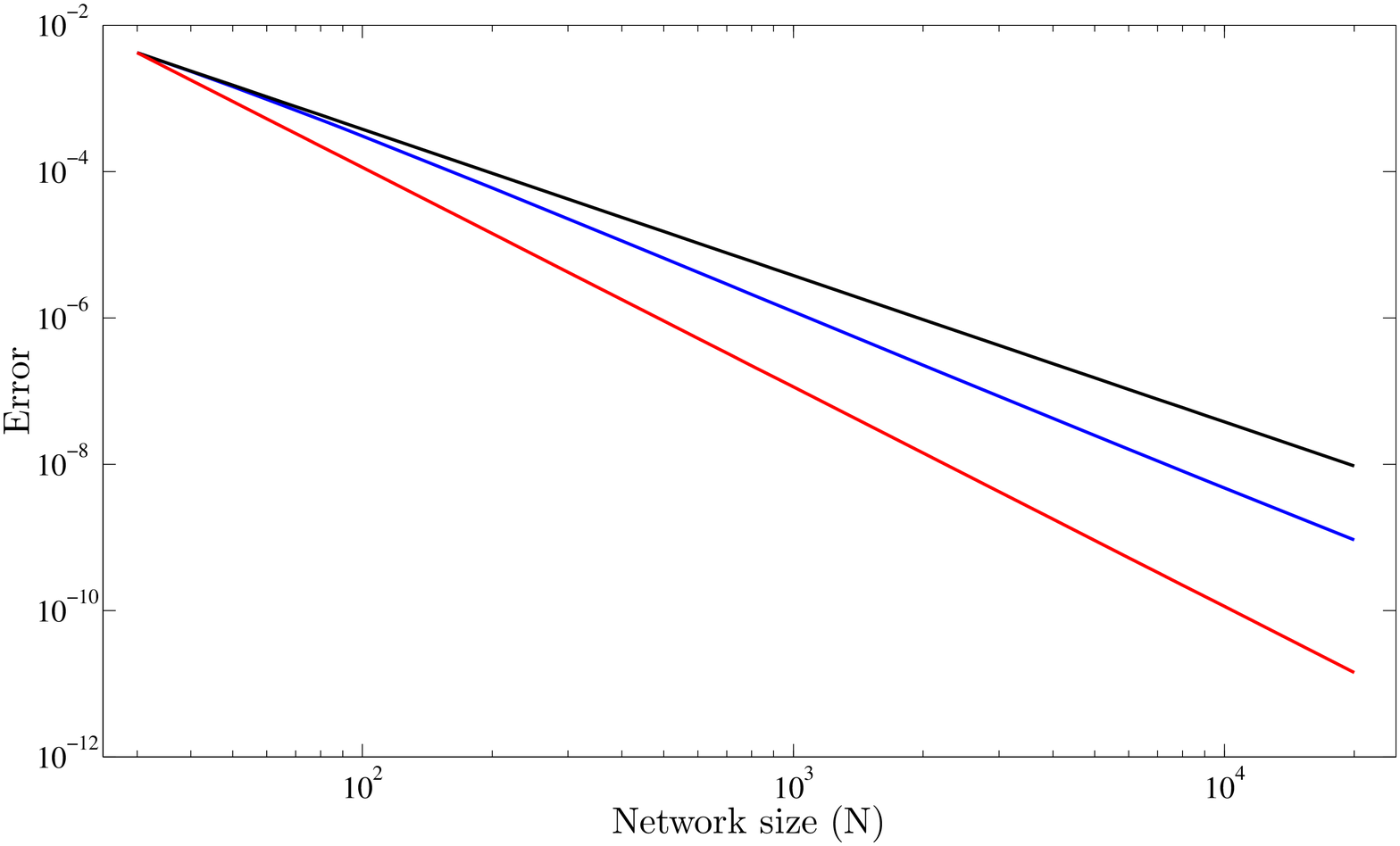}
        }\\
    \end{center}
    \caption{{\bf Convergence of closed solution to exact.} (a) Here the mean square error is given by the blue line and O($N^2$) and O($N^3$) convergence represented by the black and red lines respectively.  (b) Here the supremum error is given by the blue line and O($N^4$) and O($N^5$) convergence represented by the black and red lines respectively.}
    \label{fig5}
\end{figure}
\section*{Scale-free behaviour in the $R_0=1$ regime}
Whilst Equation~\ref{eq1} gives a power law, this equation only holds over a limited range. Equation~\ref{eqkessler}, in turn, is neither a power law nor a truncated power law. 
Here, we assess the range over which the distribution of sizes can be said to exhibit scale-free behaviour. For a rigorous assessment of this range, we employ a subset of the model selection approach described by Clauset and colleagues~\cite{clauset}. Specifically, we consider $100,000$ of the simulated avalanches described earlier and fit a truncated power law distribution of the form $P(x) = Cx^{-\alpha}$ to avalanches up to size $x_{max}=\frac{9}{10}N$ (the choice of this upper bound will be justified in the following section) by using the maximum likelihood method (here $C$ is a 
normalising constant to keep the sum of the distribution between $[x_{min},x_{max}]$ equal to $1$).  We do this by finding values of $\alpha$ and $x_{min}$ that maximise the probability of obtaining our simulated avalanches given the fitted distribution. Next we randomly generate $1000$ data sets from the fitted distribution and compute the difference between these synthetic data sets and the fitted form (using the Kolmogorov-Smirnov statistic).  Similarly we compute the difference between our simulations and the fitted power law.  The p-value is then calculated as the proportion of synthetic data sets that are further away from the theoretical distributions than our simulations.  As per ~\cite{clauset}, the  hypothesis (that the data comes from a power law) is rejected if the p-value is less than 0.1. Note that in the model selection approach, should the hypothesis not be rejected, then one should test alternative models and use an information criterion to identify the best model. However, our focus here is purely on assessing whether our distribution can be said to behave like a power law distribution (we know it is not actually a power law) and therefore alternative models were not tested. With $100,000$ avalanches we obtained a p-value of $0.382$ leading us not to reject the hypothesis that the distribution was power law (see figure~\ref{fig6}). Since the distribution is not a power law, we would expect that upon considering a larger number of avalanches, this hypothesis should be rejected~\cite{klaus}. Indeed, using data from $1,000,000$ avalanches yielded a p-value of $0$, i.e., the truncated power law is not an appropriate model for the distribution.  
 
\begin{figure}[!ht]
    \begin{center}
        \subfigure[PDF]{
            \label{fig6a}
            \includegraphics[width=0.45\textwidth]{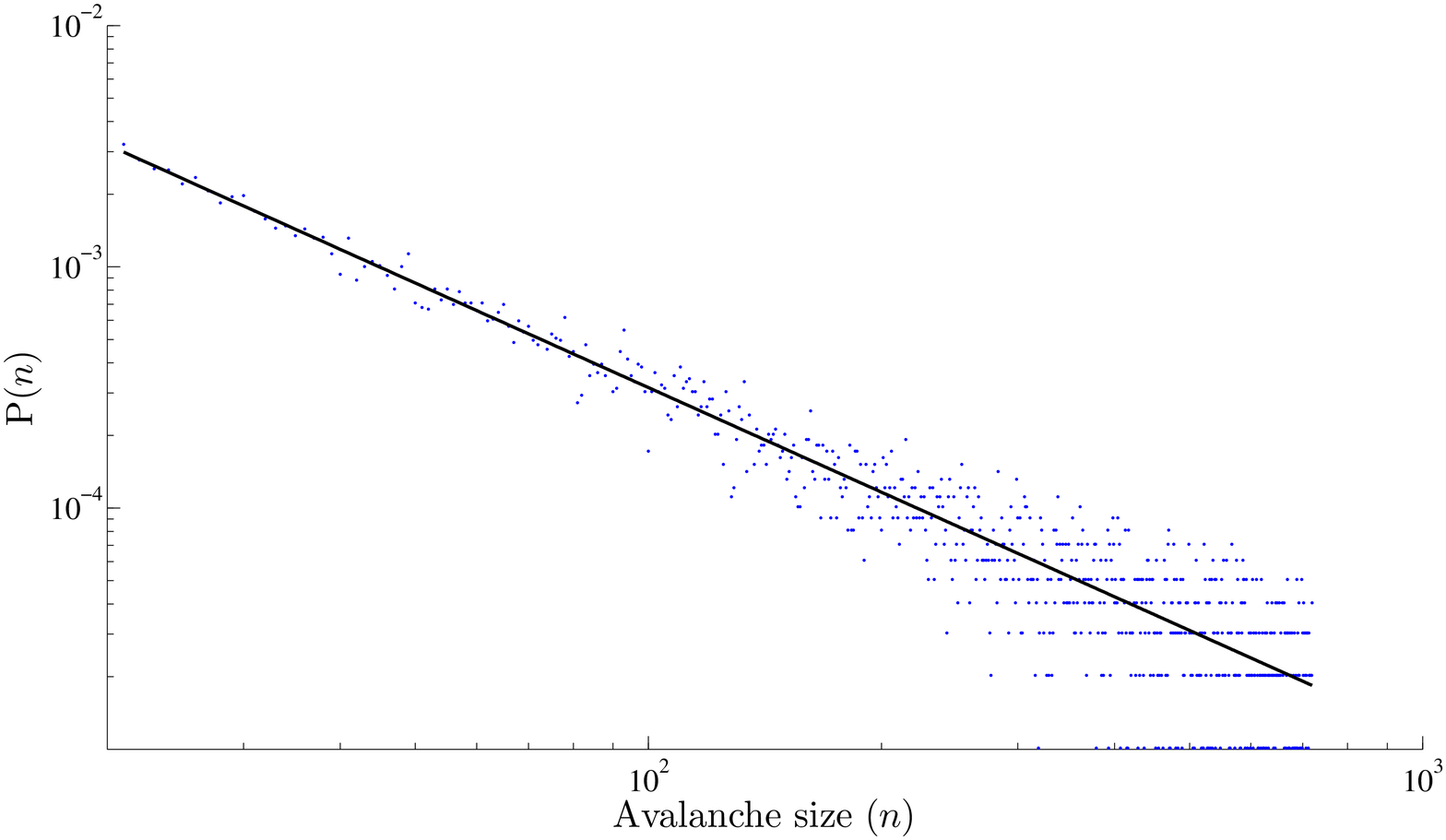}
        }
        \subfigure[CDF]{
           \label{fig6b}
           \includegraphics[width=0.45\textwidth]{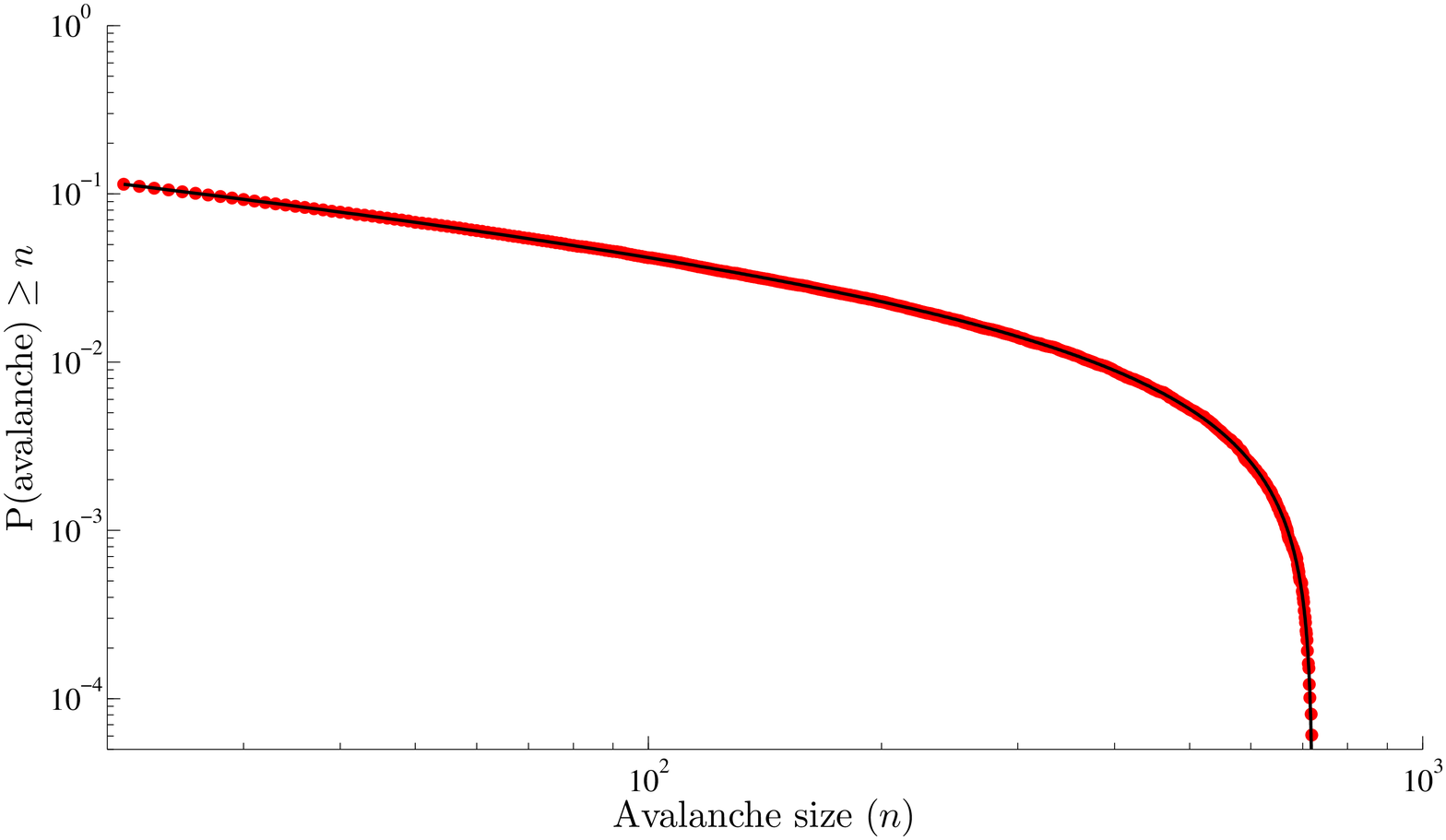}
        }\\
    \end{center}
    \caption{{\bf Fitted distributions.} Out of $100,000$ of the observed avalanches we fit the $98,833$ whose size was less than $\frac{9}{10}N$. (a) The fitted probability distribution function (black line) fitted over the simulated avalanche distribution (blue dots). (b) The fitted cumulative distribution function (black line) fitted over the simulated avalanche distribution (red dots).}
    \label{fig6}
\end{figure}

The fact that the truncated power law was a plausible fit for the lesser number of avalanches (note that $100,000$ is of the same order of magnitude as the number of avalanches typically reported in {\it in vitro} or {\it in vivo} studies of neuronal avalanches) is indicative of the partial scale-free behaviour the model exhibits. The appeal of the concept of critical brain is that the critical regime is the one in which long-range correlations keep the system poised between too highly correlated states of no behavioural value and too weakly correlated states that prevent information flow~\cite{chialvo2004}. Thus, the actual nature of the distribution of the avalanche size matters less than any indication of the presence of long range correlations. In other words, neuronal avalanches need not precisely follow a power law, they just need to exhibit similar behaviour. It is important to appreciate this distinction. As the exact solution to the distribution of avalanche sizes is known, we can then compare it visually with a fit of a truncated power law over avalanche sizes from $\frac{1}{10}N$ to $\frac{9}{10}N$.  This is done in Figure~\ref{fig7} which confirms that over a limited range of sizes the distribution is well approximated by a power law. 
\begin{figure}[!ht]
		\begin{center}
				\includegraphics[width=0.7\textwidth]{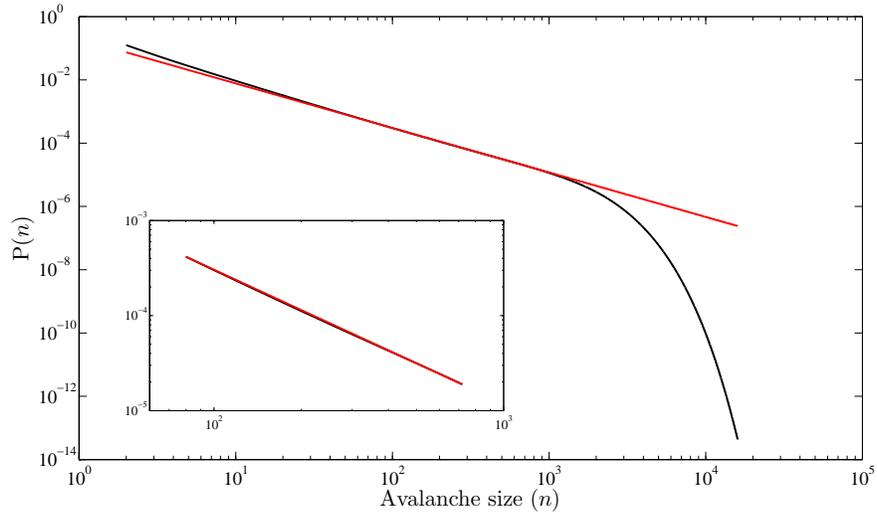}
		\end{center}
		\caption{
			{\bf Power law fit of the exact solution.} Main: plot of the truncated power law (red) fitted over the entire range of the exact distribution (black). Inset: Fitted power law and exact distribution in the range [$\frac{1}{10}N,\frac{9}{10}N$]. 			}
\label{fig7}
\end{figure}
\section*{Origin of the distribution's truncation}
The fact that we have an exact form for the distribution allows us to make further important observations about some of its characteristics. Here, we explore the origin of the distribution's truncation. Let $\lambda_{0}$, $\lambda_{1}$, $\ldots \ldots$, $\lambda_{\tilde{N}}$ be the eigenvalues of $\mathbf{A}$ with the corresponding eigenvectors $u_{0}$, $u_{1}$, $\ldots \ldots$, $u_{\tilde{N}}$.  The initial condition can then be given as $\mathbf{r}(0) = c_{0}u_{0} + c_{1}u_{1} + \ldots \ldots + c_{\tilde{N}}u_{\tilde{N}}$.  Using the property $\mathbf{A}u_{j} = \lambda_{j}u_{j}$ we then obtain $\mathbf{r}(k) = c_{0}\lambda^{k}_{0}u_{0} + c_{1}\lambda^{k}_{1}u_{1} + \ldots \ldots + c_{\tilde{N}}\lambda^{k}_{\tilde{N}}u_{\tilde{N}} $. This calculation leads to the probability of an avalanche being of size $n$ being: 
\begin{equation}
P(n)=q_1\sum_{i=1}^{\tilde{N}}d_{i}\lambda_i^n,
\label{DistP(n)}
\end{equation} 
where $q_1$ is the probability that the next transition is a recovery (from $A$ to $Q$) given 1 active neurone (as defined earlier), $\lambda_{i}$s are the eigenvalues of the transition matrix $\mathbf{A}$ and $d_i$s are specified by the eigenvectors of the transition matrix and the initial conditions. We note that the earlier equation, $\mathbf{r}(0) = c_{0}u_{0} + c_{1}u_{1} + \ldots \ldots + c_{\tilde{N}}u_{\tilde{N}}$, can be solved to obtain $c_{i}$.  Using this, we can then calculate $d_{i}$ as the first entry of the vector $c_{i}u_{i}$. 
Equation~\ref{DistP(n)}, which is exact, thus demonstrates that the distribution of avalanche sizes is a linear combination of exponentials. 
\\ \\
Assuming the lead eigenvalue is denoted by $\lambda_1$, then for all $i$, $\lambda_{i} < \lambda_1$ and we have
\begin{align*} 
\frac{P(n)}{q_1d_1\lambda_1^n}
		&=\sum_{i=1}^{\tilde{N}}\frac{d_{i}\lambda_i^n}{d_1\lambda_1^n} \\
		&=\frac{d_1}{d_1}\left(\frac{\lambda_1}{\lambda_1}\right)^n + \frac{d_2}{d_1}\left(\frac{\lambda_2}{\lambda_1}\right)^n +\ldots+ \frac{d_{\tilde{N}}}{d_1}\left(\frac{\lambda_{\tilde{N}}}{\lambda_1}\right)^n \\
\end{align*}
Taking the limit as $n$ increases we find
\begin{align*} 
		\lim_{n\rightarrow\infty} \frac{P(n)}{q_1d_1\lambda_1^n} &=1
\end{align*}
Hence $P(n) \sim q_1d_1\lambda_1^n$ and for larger avalanche sizes we have the leading eigenvalue dominating thus giving the exponential cutoff
observed.  We illustrate this convergence in Figure~\ref{fig8} where we plot the exact avalanche distribution, $P(n)$, against $q_1d_1\lambda_1^n$.  
This figure also illustrates that the leading eigenvalue begins to dominate for avalanches just over the system size.
It is for this reason that we chose an upper bound of $\frac{9N}{10}$ when fitting a power law to the distribution of avalanche sizes in the previous section. 

\begin{figure}[!ht]
		\begin{center}
				\includegraphics[width=0.7\textwidth]{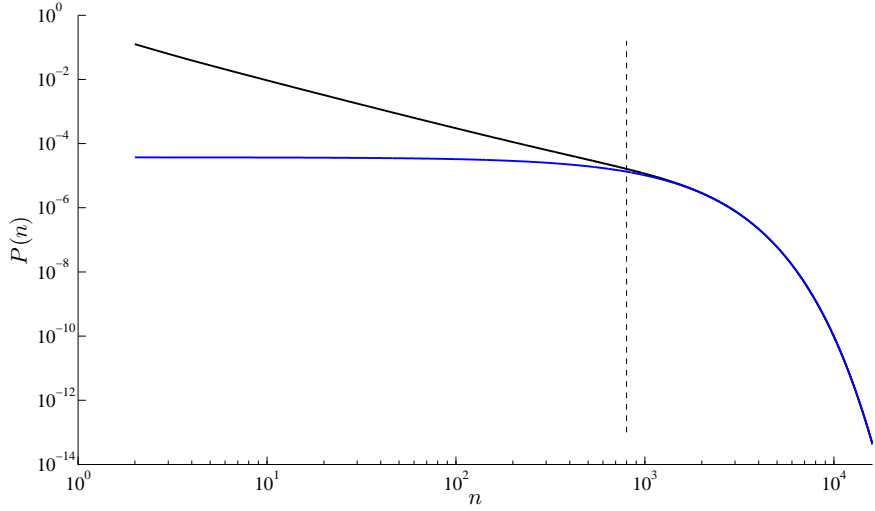}
		\end{center}
		\caption{
			{\bf Exponential cutoff.} Exact avalanche distribution (black line), plotted against a distribution assuming only the leading eigenvalue
			is non-zero (blue line).  Avalanches greater than the system size, $N=800$, appear after the dashed line.
			}
\label{fig8}
\end{figure}
\section*{Power law distribution for large systems}
Whilst we are unable to show analytically that in the limit of the system size the distribution converges to a power law, we can
conjecture that this is the case.  We motivate this by considering a hypothetical system that can again be characterised by its
transition matrix and Equation~\ref{DistP(n)} for the probability distribution.  Such a distribution could converge to a true power 
law under two important conditions: 
\begin{enumerate}
\item the eigenvalues of the transition matrix $\mathbf{A}$ are well approximated by a geometric distribution, i.e. they are in the form $\lambda_i=Ke^{-\mu i}$,
\item the constants $d_i$ in Equation~\ref{DistP(n)} are well approximated by $d_{i}=Li^q$, 
\end{enumerate}
\noindent where $K$, $\mu$, $L$ and $q$ can be inferred via a numerical fitting procedure. In such a scenario, Equation~\ref{DistP(n)} can be rewritten to give
\begin{equation}
P(n)=C\sum_{i=1}^{\tilde{N}}i^q(e^{\mu n})^{-i},
\end{equation} 
where $C$ is a given constant. In the limit of an infinite network size we then have
\begin{equation}
P(n)=C\sum_{i=1}^{\infty}i^q(e^{\mu n})^{-i}.
\end{equation}   
While $P(n)$ can be found based on standard mathematical arguments, we have chosen to use 
results derived in the context of the $\mathbf{Z}$-transform. The standard results for integer values of $q$ give
\begin{equation}
\sum_{i=1}^{\infty}i^qz^{-i}=(-1)^qD^q(\frac{z}{z-1}),  
\end{equation} 
where $D$ is an operator such that $D(f(z))=z\frac{d(f(z))}{dz}$. For a fixed integer value of $q$, an approximation for $P(n)$ can be obtained by simply applying the operator as many times as necessary and then substituting $z=e^{\mu n}$. For $q=1$ for example, $P(n) \propto \frac{e^{\mu n}}{(e^{\mu n}-1)^2}$ which for small values of $\mu$ is well approximated by $\frac{1}{\mu^2}\frac{1}{n^2}$. 

These results only hold for integer values of $q$ so an alternative approach is to approximate the sum for $P(n)$ in terms of an integral. 
Taking into account the special form for the eigenvalues and constants, $P(n)$ can be approximated as follows:
\begin{equation}
P(n)=C\sum_{i=1}^{\infty}i^q(e^{\mu n})^{-i} \simeq C\int_{0}^{\infty}x^q e^{-\mu n x}dx.
\end{equation} 
The latter integral can be interpreted as a Laplace transform of $x^q$ and thus yields
\begin{equation}
P(n) \simeq C\frac{\Gamma (q+1)}{\mu^{q+1}}\frac{1}{n^{q+1}}.
\end{equation}
It is worth noting that this result is consistent with that obtained for integer values of $q$.  

For a simple empirical verification of this conjecture, we determined the values of $K$, $\mu$, $L$ and $q$ in the above conditions that fitted the first and twentieth eigenvalues and $d$ constants of the exact distribution for a network of size $N=800$ (see figures~\ref{fig9a}~and~\ref{fig9b}) and compared the resulting probability distribution with the exact distribution. As shown by figure~\ref{fig9c}, there is remarkable agreement between both curves over a broad range of values, including the range  [$\frac{1}{10}N,\frac{9}{10}N$] over which a power law like behaviour was established earlier (see figure~\ref{fig7}). This result clearly illustrates the dominance of the larger eigenvalues and, given that the fitted distribution converges to a power law, gives support to the conjecture that the exact distribution would do so in the limit of an infinite network.

\begin{figure}[!ht]
    \begin{center}
        \subfigure[distribution of eigenvalues $\lambda_i$]{
            \label{fig9a}
            \includegraphics[width=0.45\textwidth]{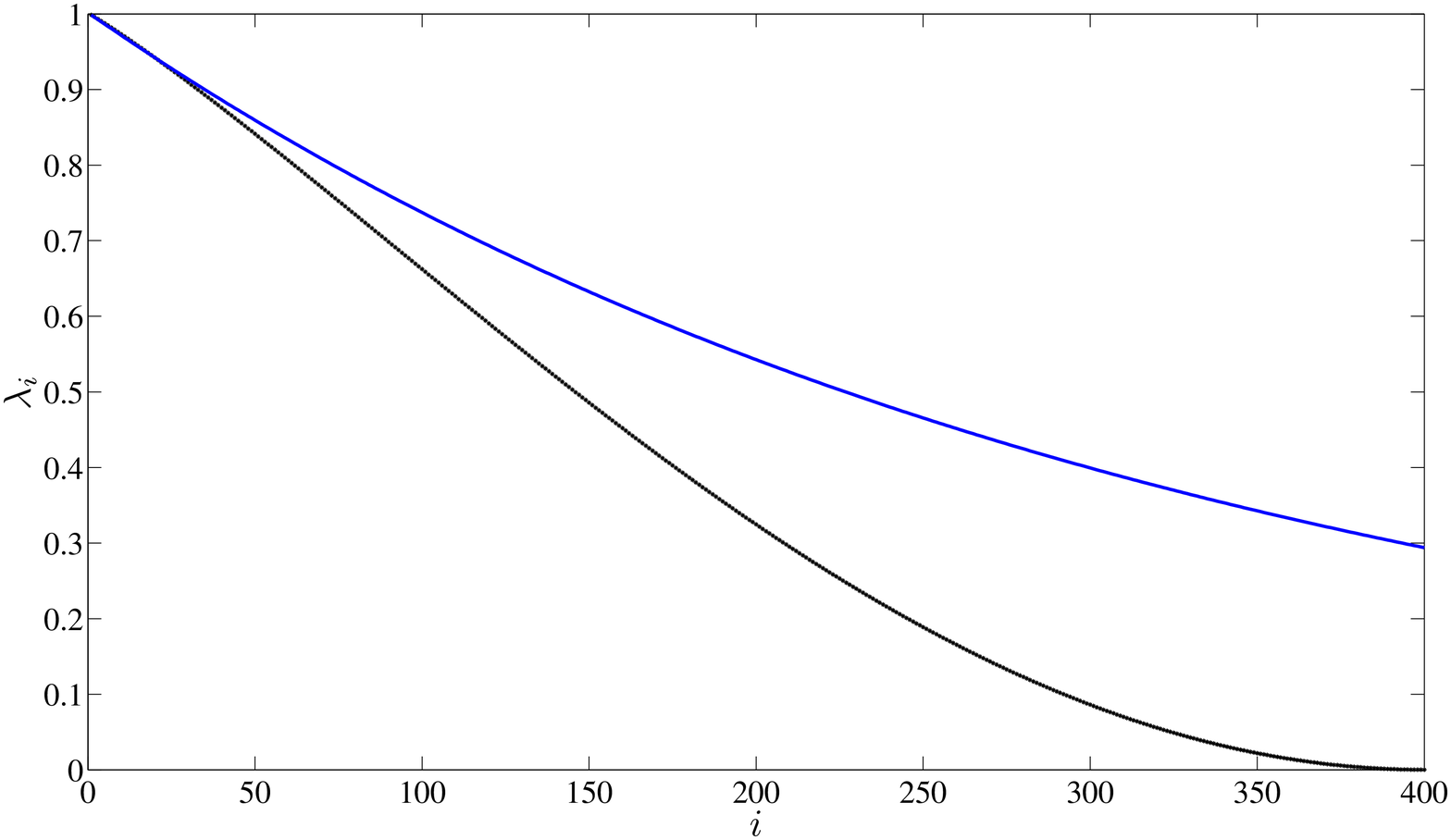}
        }
        \subfigure[distribution of constants $d_i$]{
           \label{fig9b}
           \includegraphics[width=0.45\textwidth]{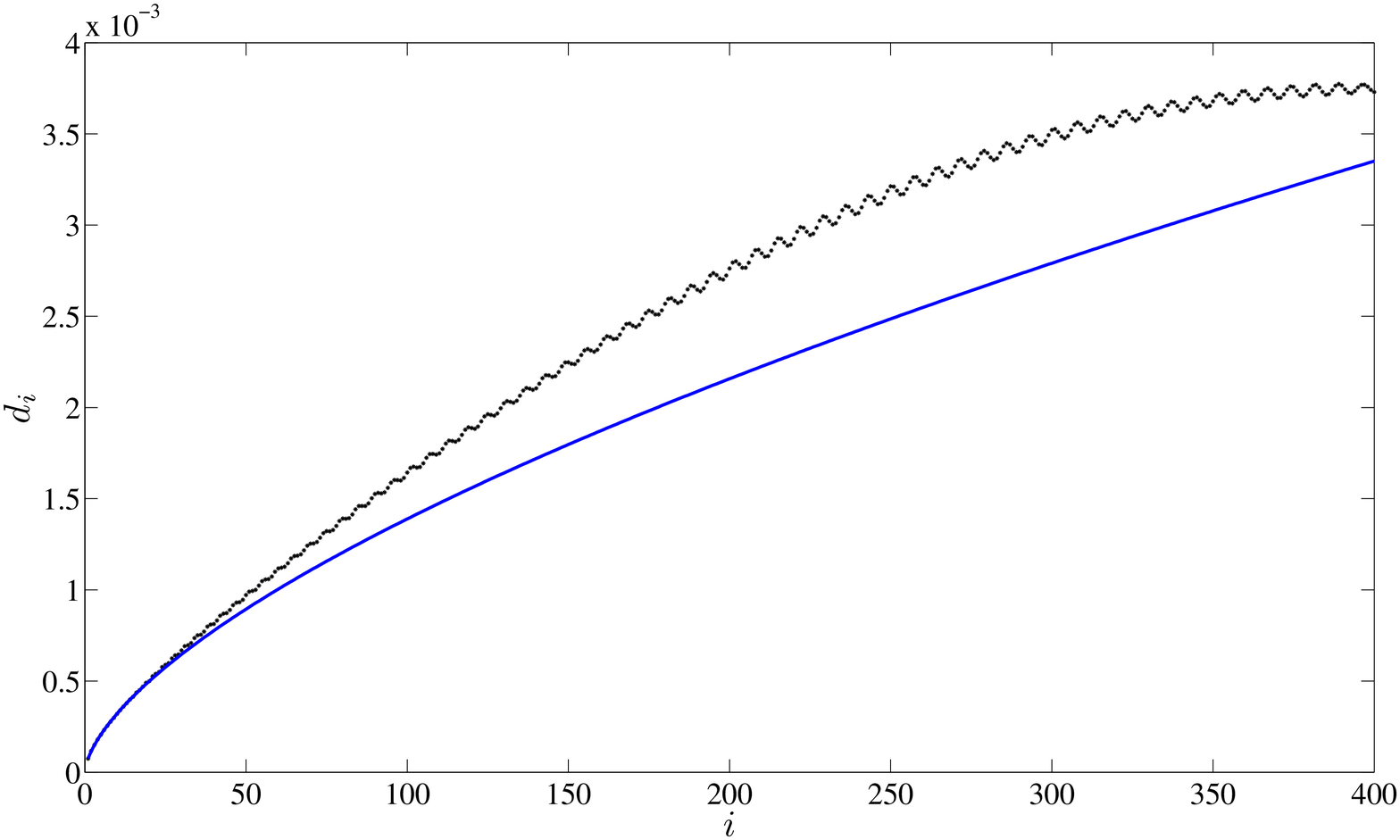}
        }\\
        \subfigure[distribution of avalanche sizes]{
           \label{fig9c}
           \includegraphics[width=0.45\textwidth]{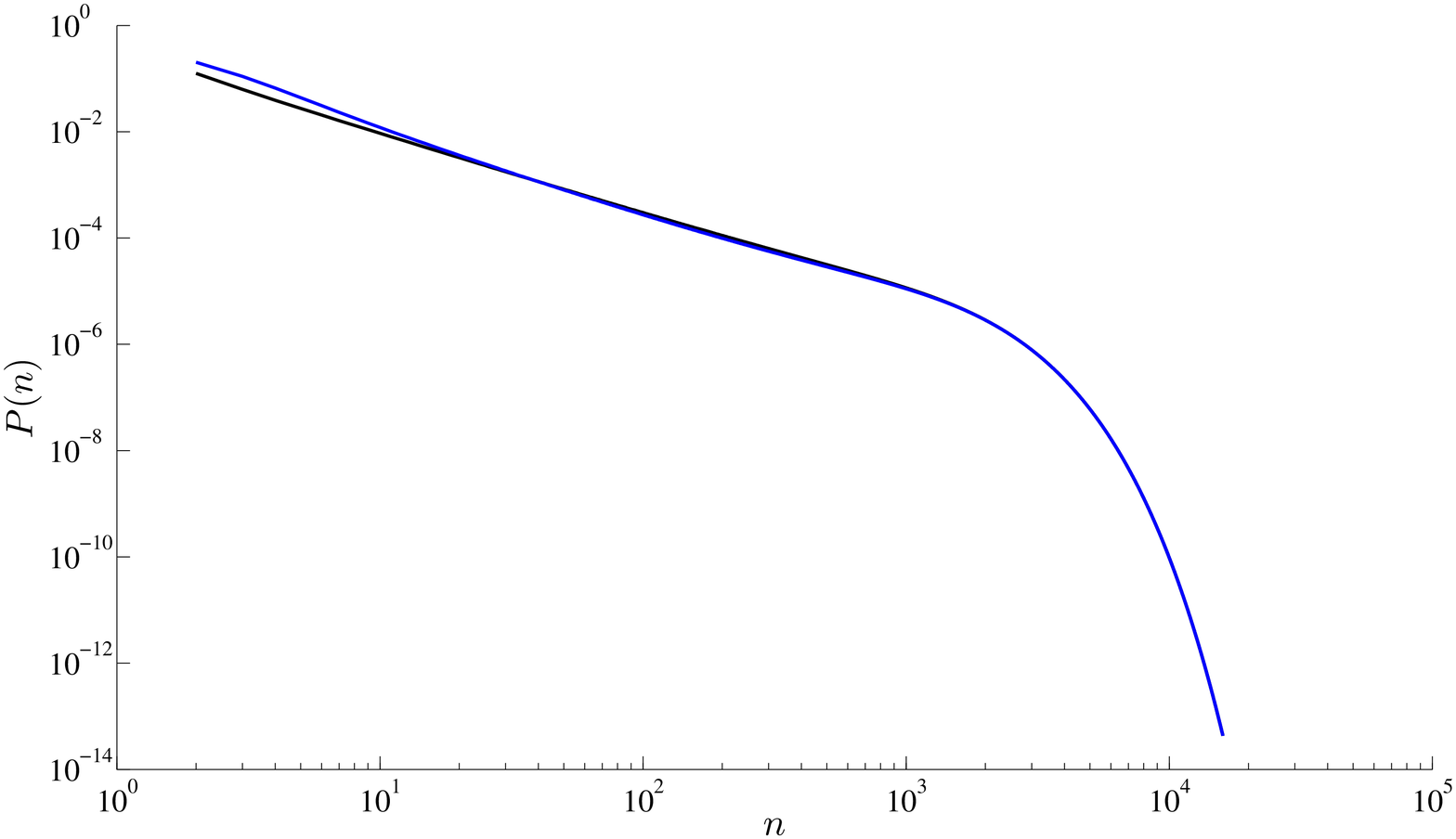}
        }   
    \end{center}
    \caption{{\bf Possible origin of the power law for large systems.}  
    (a) Actual distribution of eigenvalues $\lambda_i$ (black crosses) along with fitted distribution (blue line).  
    (b) Actual distribution of constants $d_i$ (black crosses) along with fitted distribution (blue line).  
    (c) Exact distribution of avalanche sizes (black) along with distribution resulting from fitted distributions of $\lambda_i$ and $d_i$ (blue). 
    All plots are for a network of size $N = 800$ operating at criticality.}
    \label{fig9}
\end{figure}
\section*{Other markers of criticality}
Since the distribution of avalanche sizes in the finite-size critical system does not necessarily follow a true power law, the application of robust statistical testing in experimental conditions could well lead to rejecting the hypothesis that the data may come from a system operating in the critical regime. Therefore, in this section, we consider two experimentally testable markers of criticality: critical slowing down and divergence of susceptibility. We will define those concepts below  but first we briefly summarise Van Kampen's system size expansion~\cite{vankampen} which we use to illustrate those markers on our system.

\subsection*{System size expansion}
For generality we now assume that each neurone receives a constant external input and that the activation function can take forms other 
than the simple identity function. We define the probability that the number of neurones active at time $t$ is $n$ as $P_n(t)$.  Then the master
equation can be given as
\begin{align*}
\frac{d P_{n}\left(t\right)}{d t} =  	&\alpha\left(n+1\right)P_{n+1}\left(t\right) - \\ 
							&\alpha nP_{n}\left(t\right) + \\ 
							&f\left(\frac{w(n-1)}{N} + h\right)\left(N-\left(n-1\right)\right)P_{n-1}\left(t\right) -  \\ 
							&f\left(\frac{wn}{N} + h\right)\left(N-n\right)P_{n}\left(t\right).
\end{align*}
The idea is to now model the number of active neurones as the sum of a deterministic component scaled by $N$ and a stochastic perturbation scaled by $\sqrt{N}$, i.e., 
\begin{align*}
n(t) &= N\mu(t) + N^{\frac{1}{2}}\xi(t).
\end{align*}
Full details of the system size expansion are given in the Appendix, but what we obtain is the following set of equations for $\mu$ (which is the solution to the mean field equation of the proportion of active neurones), $\left<\xi\right>$ (the expected value of the 
fluctuations) and $\sigma^2 = \left<\xi^2\right> - \left<\xi\right>^2$ (the variance of the fluctuations)
\begin{align}
\frac{\partial\mu}{\partial t} &= -\alpha\mu + (1-\mu)\hat{f}\label{mueq},\\
\frac{\partial \left<\xi\right>}{\partial t} &= -\left(\alpha + \hat{f} - w\hat{f}'(1-\mu)\right)\left<\xi\right>\label{xieq}, \\
\frac{\partial \left<\sigma^2\right>}{\partial t} &= -2\left(\alpha + \hat{f} - w\hat{f}'(1-\mu)\right)\left<\sigma^2\right>
+\left(\alpha\mu + (1-\mu)\hat{f}\right). \label{syssizeexp}
\end{align}
Here $\hat{f}=f(w\mu +h)$ and $\hat{f}'=f'(w\mu +h)$.  These equations, in turn, give the following equations for the 
mean, $A$, and variance, $A_\sigma$, of the number of active neurones
\begin{align}
A &= N\mu + N^{-\frac{1}{2}}\left<\xi\right> = N\mu \quad \text{(assuming we know the initial number of active neurones)}\label{system_mu}, \\
A_{\sigma} &= N\left<\sigma^2\right>\label{system_var}.
\end{align}
\subsubsection*{Critical slowing down}
In dynamical systems theory, a number of bifurcations, including the transcritical bifurcation in our system, involve the dominant eigenvalue characterising the rates of changes around the equilibrium crossing zero. As a consequence, the characteristic return time to the equilibrium following a perturbation increases when the threshold is approached~\cite{wissel}. This increases has led to the notion of critical slowing down as a marker of critical transitions~\cite{scheffer}. 
Here, we illustrate the critical slowing down of our model with the analytic derivation of the rate of convergence to the steady state (this derivation has been previously shown by~\cite{stollenwerk}). 
We first begin by calculating the analytic solution to Equation~\ref{mueq} for the percentage of active neurones.  
We again consider the case where $f$ is the identity function and can thus write      
\begin{align}
		\frac{\partial \mu}{\partial t} 
				&= -\alpha\mu + (1-\mu)f(w\mu + h) = -\alpha\mu + (1-\mu)(w\mu + h).
				\label{mueq2}
\end{align}
Assuming zero external input ($h = 0$),  we have
\begin{align}
		\frac{\partial \mu}{\partial t} 
				&= -\alpha\mu + (1-\mu)(w\mu+h) = -\alpha\mu + (1-\mu)w\mu.
\end{align}
We are interested in the solution of this equation and consider the result for different values of $\alpha$.  
Firstly we consider $\alpha \neq w$.  In this case we have
\begin{align} 
		\frac{\partial \mu}{\partial t}
				&= -\alpha\mu + (1-\mu)w\mu = \mu(w-w\mu-\alpha).
\end{align}
Integrating this using separation of variables and the initial condition $\mu(0) = \mu_{0}$, we find
\begin{align}
	\mu(t) &= \frac{w-\alpha}{Ae^{(\alpha-w)t}+w} \quad \text{ where } A = \frac{\mu_0}{w-w\mu_0 - \alpha}.  
\end{align}
The solution to this depends on whether $\alpha<w$ or $\alpha>w$ ($R_0 > 1$ and $R_0 < 1$ respectively).  If $\alpha<w$ then as 
$t\rightarrow \infty$, $\mu \rightarrow \frac{w-\alpha}{w}$.  If $\alpha>w$ then as $t\rightarrow \infty$, $\mu \rightarrow 0$. Note that in both cases, convergence of the number of active neurones to the steady state solution is exponential. 

Now we consider the solution when $\alpha = w$, i.e., the critical regime.
\begin{align*}
		\frac{\partial \mu}{\partial t}
				&=-\alpha\mu + (1-\mu)\alpha\mu =-\alpha\mu^2 \Rightarrow \mu(t) = \frac{1}{\alpha t+\mu_{0}^{-1}}.
\end{align*}
Thus as $t\rightarrow\infty$ we find $\mu(t)\rightarrow 0$. However,  unlike for $R_{0} \neq 1$, convergence to the steady state exhibits a power law dependence on time~\cite{stollenwerk}. 


\subsubsection*{Divergence of susceptibility}
A correlate of the phenomenon of critical slowing down is that of the divergence of susceptibility of the system as the system approaches the bifurcation~\cite{scheffer}. In this section, we investigate the behaviour of the equation for the variance. For simplicity, we consider again the case of the identity activation function and a non-driven system $h=0$. First we use Equation~\ref{syssizeexp} to calculate the variance in the percentage of active neurones: 
\begin{align*}
		\frac{\partial\sigma^2}{\partial t}
			&= -2\left(\alpha + \hat{f} - w\hat{f}'\left(1-\mu\right)\right)\sigma^2 + \left(\alpha\mu + \left(1-\mu\right)\hat{f}\right) \\
			&= -2\left(\alpha + w\mu+h - w^2\left(1-\mu\right)\right)\sigma^2 + \left(\alpha\mu + \left(1-\mu\right)\left(w\mu+h\right)\right) \\
			&= -2\left(\alpha + w\mu - w^2\left(1-\mu\right)\right)\sigma^2 + \left(\alpha\mu + \left(1-\mu\right)w\mu\right) 
\end{align*}				
Setting this equal to zero and rearranging we obtain
\begin{align*}
		\sigma^2 
			&= \frac{\left(\alpha\mu + \left(1-\mu\right)w\mu\right)}{2\left(\alpha + w\mu - w^2\left(1-\mu\right)\right)} 
			= \frac{\left(\mu + \left(1-\mu\right)R_0\mu\right)}{2\left(1 + R_0\mu - R_0w\left(1-\mu\right)\right)}. 		
\end{align*}
Here we note that unlike the equation for $\mu$ where there was only the single bifurcation parameter $R_0$, we now have the additional dependence on $w$.  To maintain consistency with earlier results, we now set $w=1$ to obtain
\begin{align*}
\lim\limits_{t \rightarrow \infty}\sigma^2(t) = 
	\begin{cases} 
		\alpha &\mbox{if } \alpha < 1 \quad (R_0 > 1), \\
		\frac{1}{2} &\mbox{if } \alpha = 1 \quad (R_0 = 1), \\
		0 & \mbox{otherwise } \quad (R_0 < 1).\\ 
	\end{cases}. 
\end{align*}
Using Equation~\ref{system_var} we obtain 
\begin{align*}
\lim\limits_{t \rightarrow \infty}\left<A\right>_{\sigma} = \lim\limits_{t \rightarrow \infty}N\left<\sigma^2\right> = 
	\begin{cases} 
		\frac{N}{R_0} &\mbox{if } R_0 > 1, \\
		\frac{N}{2} &\mbox{if } R_0 = 1, \\
		0 & \mbox{otherwise } (R_0 < 1).\\ 
	\end{cases}
\end{align*}
Figure~\ref{fig10} illustrates the jump to a non-zero steady state when the critical value $R_0=1$ is approached from below, and the divergence in variance when it is approached from above. 
\begin{figure}[!ht]
		\begin{center}
				\includegraphics[width=0.7\textwidth]{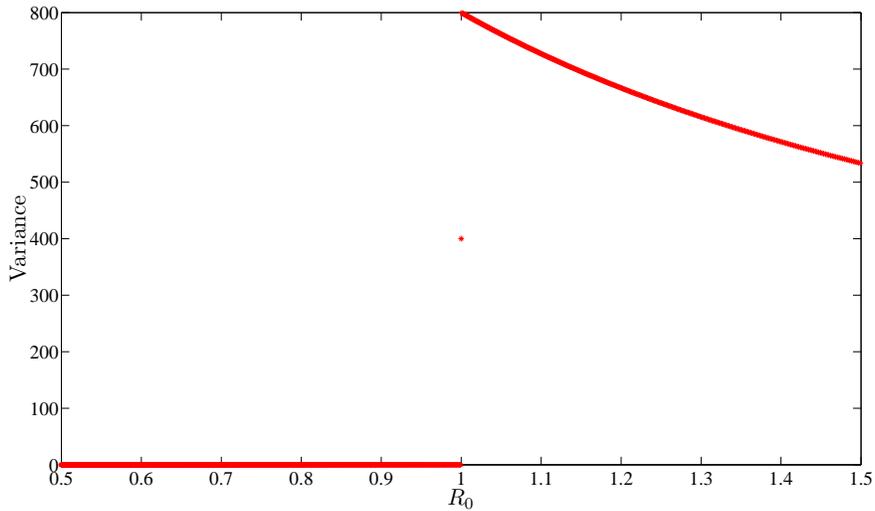}
		\end{center}
		\caption{
			{\bf Divergence of susceptibility.} Analytic result for the steady state of the variance as $R_0$ approaches $1$
			in a network of size $N = 800$.  Results only provided down to $\alpha=2/3$ for clarity.}
\label{fig10}
\end{figure}

Here it should be noted that any finite-size network has a zero absorbing state so that eventually all activity will die out irrespective of the value of $R_0$.  However, it has been shown that the ODE limit is a valid approximation to the solution of the master equation for reasonably sized systems with values of $R_0$ greater than $1$ and only over a finite time horizon (see \cite{nasell} for further discussion). Defining the true (i.e., calculated directly from the master equation for $P(n)$) expected value of active neurones at time $t$ as $\tilde{A}(t)$, the convergence of the ODE approximation for $A(t)$ given by Equation~\ref{system_mu} is such that for any $t \geq 0$ $\lim\limits_{N \rightarrow \infty} |A(t)-\tilde{A}(t)|=0$~\cite{simon}.

\section*{Discussion}
Over the last decade or so, the search for evidence that the brain may be a critical system has been the focus of much research. This is because it is thought that a critical brain would benefit from maximised dynamic range of processing, fidelity of information transmission and information capacity~\cite{shew2012}. Whilst support for the critical brain hypothesis has emerged from comparing brain dynamics at various scales with the dynamics of physical systems at criticality (e.g.,~\cite{plenzandchialvo,expert,linkenkaer2001,poil2012,friedman,ribeiro}), in this paper, we focus on the important body of work that has relied on characterising power laws in the distributions of size of neuronal avalanches~\cite{beggsorg,shew}. Our focus on this scale is motivated by empirical considerations regarding how one can go about demonstrating the above functional properties. Shew and Plenz~\cite{shew2012} remark that any research strategy to test whether these properties are optimal near criticality will have to achieve two criteria: a means of altering the overall balance of interactions between neurones and a means of assessing how close to criticality the cortex is operating. As argued by these authors, the study of neuronal avalanches offers the greatest likelihood of achieving those two criteria. 

The importance of a robust assessment of the statistical properties of the avalanche size is therefore two-fold: on the one hand, it is about ascertaining the extent to which the system being studied has the statistical properties expected of a system operating at, or near, criticality; on the other hand, it is about being able to confirm that a manipulation/perturbation of the system aimed to push the system away from this critical regime has been effective. This consideration therefore puts a lot of importance on the description of the statistics one should expect in such a system. In the current literature, the assumption of the distribution of avalanche sizes taking a power law functional form relies on an analogy between the propagation of spikes in a neuronal network and models of percolation dynamics or branching processes for which exact power laws have been demonstrated {\em in the limit of system size}. As a result of the importance of having a robust assessment of the expected presence of a power law, greater emphasis has recently been put on using a sound statistical testing framework, e.g.,~\cite{clauset}. Whilst we are unaware of any study in which the criticality hypothesis was rejected due to failure of rigorous statistical testing (which we suspect is due to the necessarily small number of observations, as we will argue below), there is clear evidence that many authors are now using the methods of Clauset et al.~\cite{clauset} to confirm the criticality of their experimental findings, e.g.,~\cite{klaus,touboul,benayoun}. As a result, we feel that it is all the more important to confirm that the assumed power law functional form is indeed a sensible representation of what one should expect in {\it in vivo} and {\it in vitro} recordings, which, unlike the physical systems considered when deriving the power law statistics, are finite-size systems. The aim of the paper was therefore to consider a model of neuronal dynamics that would be simple enough to allow the derivation of analytical or semi-analytical results whilst (i) giving us a handle on the parameter controlling the fundamental principle thought to underlie criticality in the brain, namely, the {\em balancing} between processes that enhance and suppress activity (note that we are intentionally not referring to excitation and/or inhibition -- we will return to this below) and (ii) allowing us to determine its distribution of avalanche sizes when operating in the critical regime. Note that because we are using a finite-size system, we are appealing to a normal form of standard bifurcation, here, a transcritical bifurcation, because it embodies all that needs to be known about the `critical' transition (Sornette, private communication). 

Our semi-analytic derivation of the true distribution of avalanche sizes in a finite-size system suggests that, even though it is approximately scale free over a limited range, the distribution is not a true power law. First, this has important implications for the interpretation of results from a robust statistical assessment of the distribution. Indeed, as has been discussed by Klaus and Plenz~\cite{klaus}, with a large number of samples, any distribution that deviates from the expected distribution by more than noise due to sampling, will eventually yield a p-value such that the power law hypothesis will be rejected, thus leading to the potentially incorrect conclusion that the system is not critical. This is the case in our scenario where using $10^6$ avalanches lead to a rejection of the criticality hypothesis even though the system is tuned to the critical regime. In contrast, with $10^5$ avalanches (which is consistent with empirical observations), a p-value above threshold leads to not rejecting the hypothesis that the distribution is a power law even though we established it is not one\endnote{As the power law is not a sufficient condition of criticality, one should not infer from this that the system is indeed critical, however, this step is commonly taken in published reports and that is worth mentioning here.}. This finding therefore provides an important counterpart to the analytical results of Touboul and colleagues~\cite{touboul} who showed that thresholded stochastic processes could generically yield apparent power laws that only stringent statistical testing will reject. Whilst the stringent testing will reject the hypothesis of criticality for a system that is not necessarily critical, it may also reject the hypothesis of criticality for a system that is critical only because the actual distribution is not actually a power law. This ambiguity of the avalanche distribution in the finite-size system therefore requires that one should carefully consider to what fundamental property the idea of a critical brain actually appeals to. We suggest that the key appeal is that the brain can exhibit long-range correlations between neurones without it ever experiencing an over saturation of activity or long periods of inactivity. It then follows that the importance is not in the exact distribution obtained but in the approximately scale-free behaviour it exhibits. In turn, this highlights the importance of looking at other markers of criticality (which we will discuss below). 

Another important result of this work is to provide the beginning of a mechanistic explanation for an often alluded to (e.g.,~\cite{rubinov}) but never properly treated (as far as we are aware) observation that whereas avalanches in a critical system with re-entrant connections could in principle be arbitrarily long, and certainly, exceeding the number of recording sites, neuronal avalanches in {\it in vitro} or {\it in vivo} systems (and many computational models of self-organised criticality) often show a cut-off at the number of sites. Our work suggests that the lead eigenvalue of the transition matrix between states fully determine the location of this cut-off which turns out indeed to be at about the system size, even if avalanches of up to 20 times the system size can be observed. This finding therefore provides some justification for setting, or accepting, a bound within which to apply a Clauset-type methodology (we note that various reports use different ranges, e.g., 80\% of system size in~\cite{levina}, roughly system size in~\cite{rubinov}). It is worth remembering that the number of recording sites can have profound implications on the nature of the distribution observed~\cite{priesemann}. 

In addition to providing results on the distribution of avalanche sizes, we also sought to explore other potential markers of criticality. We provided results on two other markers of criticality -- critical slowing down and divergence of susceptibility -- both of which again follow from a dynamical systems appreciation of a critical bifurcation, i.e., the behaviour of a system whose lead eigenvalue crosses zero. The appeal of those markers, which have been documented in many other natural processes, e.g.,~\cite{scheffer,kelso}, but seldom at the mesoscopic brain level\endnote{Strictly speaking the notion of critical slowing in neurones firing near firing threshold appeals to the same notion.}, see~\cite{steynross} for a rare example, is that (a) they strengthen the assessment of the system being critical and (b) may contribute to achieving the second criterion of Shew and Plenz~\cite{shew2012}. Although the authors are not in a position to provide explicit recommendations for an experimental design, we believe that these markers are amenable to robust experimentation, e.g., through pharmacological manipulation. 

Whilst we hope we have convinced the reader of the potential importance of these findings, we also need to recognise that the very simplicity that makes analytical work possible does also raise questions regarding how physiologically plausible such a model is and therefore whether its conclusions should be expected to hold. Below, we address a few of the points worthy of further consideration. 

\subsection*{Validity of a purely excitatory network}
In this paper, we have used a purely excitatory neuronal model. This not only simplifies the system but is also an important characteristic of the brain during early development. Experimental results have shown that during early development, before birth, GABAergic neurones (i.e. neurones which will later be inhibitory) have a depolarising effect on their post-synaptic neighbours~\cite{cherubini,rivera,benari}. Thus, our model might be considered as representative of early development. Power law statistics have been observed in early development at a time when networks are thought to be purely excitatory~\cite{gireesh,hartley}. It should be noted that this approach has the benefit of casting a new light on the question of what is the minimum requirement for a neuronal system to show criticality. To a large extent, the current literature has been focused on a form of homeostasis resulting from either a fine balance between excitation and inhibition, e.g.,~\cite{benayoun,magnasco} or some relatively complex dynamical processes at synaptic level, e.g.,~\cite{levina}. Our results show that a purely excitatory system can show the exact same behaviour such that on average each active neurone only activates one postsynaptic neurone. Here, this balanced state is achieved through a trade-off between the rates at which neurones become active and quiescent. It should be noted that this formulation of the problem leads to interesting parallels with classical models of mathematical epidemiology which the authors intend to continue exploring. 

\subsection*{Spatial structure}
To make use of the analytic tractability of the mean field equation it was necessary to consider a fully connected network. While this is not true of the whole brain, it {\em may} be closer to the reality of the kind of in vitro systems typically considered in studies of neuronal avalanches. For example, Hellwig et al.~\cite{hellwig} report up to $80\%$ connection probability in local connectivity between pyramidal neurones in layers 2/3 of the rat visual cortex. Extending the work presented here to consider the effect of network topology on the system's dynamics and the resulting distribution of event sizes would be of particular interest from a developmental viewpoint. As networks mature, there is not only a switch to inhibition by a proportion of the neurones (the so-called GABA switch), but also a subsequent pruning of synaptic connections~ \cite{huttenlocher}. The level of pruning is high, with a $40\%$ reduction in the number of synaptic connections between early childhood and adulthood~\cite{huttenlocher}. Thus, a developing network may be more readily approximated by a fully connected network than an adult neural network would be. 

The lack of a spatial embedding of our model is in contrast with many classical models of criticality, and also with physiological systems. Accordingly, our model cannot display another important marker of criticality, namely, the divergence of correlation lengths in space. A spatial embedding is not needed for our system to be critical and to exhibit a distribution of avalanche size similar to that observed in physiological neuronal avalanches. It therefore begs the question of the exact role of spatial embedding in the dynamics of neuronal avalanches. It may well be that, just like balanced activity in our model comes about from a trade-off between excitation and refractoriness rather than between excitation and inhibition, specific spatial embeddings may enable balanced activity without the need for plastic mechanisms. Kaiser and Hilgetag~\cite{kaiser} showed that hierarchical modular networks can lead to limited sustained activity whereby the activity of neural populations in the network persists between the extremes of either quickly dying out or activating the whole network. Roxin and colleagues~\cite{roxin} observed self-sustained activity in excitable integrate-and-fire neurones in a small-world network, whose dynamics depends sensitively on the propagation velocity of the excitation. 

\subsection*{Non-driven case}
Finally, in this paper, we have focused on the non-driven case $h=0$. Whilst this constraint allowed the derivation of analytical results, it obviously contrasts with the reality of a physiological system unless one considers that any `external' input operates at such a slower timescale that one could assume separation of time scales (an important assumption in the self-organised criticality framework). However, the fact that binning is required for identifying avalanches in physiological recordings suggests that this separation of time scales is unlikely. Whilst the introduction of a non-zero $h$ in our model does not affect the results obtained using finite size expansion, it does effectively make it impossible for the system to operate at $R_0=1$. A thorough investigation of the driven case ($h>0$) will be the subject of the companion paper.

\section*{Appendix - Van Kampen's system size expansion}
For generality we now assume that each neurone receives a constant external input and that the activation function can take forms other 
than the simple identity function. If $P_n(t)$ is the probability that the number of neurones active at time $t$ is $n$, then the master
equation is given by
\begin{align*}
\frac{d P_{n}\left(t\right)}{d t} =  	&\alpha\left(n+1\right)P_{n+1}\left(t\right) - \\ 
							&\alpha nP_{n}\left(t\right) + \\ 
							&f\left(\frac{w(n-1)}{N} + h\right)\left(N-\left(n-1\right)\right)P_{n-1}\left(t\right) -  \\ 
							&f\left(\frac{wn}{N} + h\right)\left(N-n\right)P_{n}\left(t\right).
\end{align*}

Defining the step operator $\left[\varepsilon\right]$ (hereafter, in order to distinguish between operators and functions, we will use square brackets to denote operators and parentheses to denote functions, i.e., $[A](f)$ means operator $A$ acts on function $f$), 
\begin{align*}
&\left[\varepsilon\right]\left(g\left(n\right)\right) = g\left(n+1\right), \\
&\left[\varepsilon^{-1}\right]\left(g\left(n\right)\right) = g\left(n-1\right), 
\end{align*}
we can now write
\begin{align}
\frac{d P_{n}(t)}{d t} = 	&\left[\varepsilon\right]\left(\alpha n P_{n}(t)\right) - \alpha n P_{n}(t) + \left[\varepsilon^{-1}\right](f\left(\frac{wn}{N}+h\right)(N-n)P_{n}(t)) - f\left(\frac{wn}{N}+h\right)(N-n)P_{n}(t) \nonumber \\
									=& \left[(\varepsilon-1)\right](\alpha n P_{n}(t)) + \left[(\varepsilon^{-1} - 1)\right]( f\left(\frac{wn}{N}+h\right)(N-n)P_{n}(t)).
\label{dPn}
\end{align}
The following analysis is based on introducing a continuous time-dependent density function, where the discrete state space becomes continuous.  We do this by
assuming that the fluctuations about the microscopic value of $n$ are of order $N^{\frac{1}{2}}$.  In other words, we expect that $P_{n}(t)$ will have a maximum around the macroscopic value of $n$ with a width of order $N^{\frac{1}{2}}$. 
We thus take the ansatz that $n(t) = N\mu(t) + N^{\frac{1}{2}}\xi(t)$.  Here $\mu(t)$ satisfies the macroscopic (mean field) equation and $\xi$ is a stochastic perturbation scaled by $N^{\frac{1}{2}}$.  
We can now write the distribution, $P_n(t)$ as a function of $\xi$, $P_{n}(t) \approx \Pi(\xi,t)$.  Doing so allows us to approximate the moments of the master equation using PDEs.
\\ \\
Since $\left[\varepsilon\right](n) = N\mu(t) + N^{\frac{1}{2}}\xi(t)+1=N\mu + N^{\frac{1}{2}}(\xi + N^{-\frac{1}{2}})$, we can say that $\left[\varepsilon\right]$ sends $\xi \rightarrow \xi + N^{-\frac{1}{2}}$ and we get
\begin{align*}
\left[\varepsilon\right]\left(P_{n}(t)\right) &= P_{n+1}(t) \Rightarrow \left[\varepsilon\right] \left(\Pi(\xi,t)\right) = \Pi(\xi+N^{-\frac{1}{2}},t).
\end{align*}
\\ \\
Using Taylor's theorem we have
\begin{align*}
\left[\varepsilon\right]\left(f(\xi)\right) 	& = f(\xi + N^{-\frac{1}{2}}) = f(\xi) + \frac{N^{-\frac{1}{2}}}{1!}\frac{\partial f(\xi)}{\partial \xi} + \frac{N^{-1}}{2!}\frac{\partial^2 f(\xi)}{\partial \xi^2} + \ldots.
\end{align*}
Thus, we can expand the step operator $\left[\varepsilon\right]$ (and similarly $\left[\varepsilon^{-1}\right]$) as
\begin{align*}
\left[\varepsilon\right] &= I + N^{-\frac{1}{2}}\frac{\partial }{\partial \xi} + \frac{N^{-1}}{2}\frac{\partial^2}{\partial \xi^{2}} + \ldots, \\
\left[\varepsilon^{-1}\right] &= I - N^{-\frac{1}{2}}\frac{\partial }{\partial \xi} + \frac{N^{-1}}{2}\frac{\partial^2}{\partial \xi^{2}} - \ldots.
\end{align*}
Now $n = N\mu + N^{1/2} \xi \Rightarrow \xi = N^{-\frac{1}{2}}n - N^{\frac{1}{2}}\mu$.  Remembering we are interested in $P_n(t)$ for constant $n$ we then have
\begin{align*}
\frac{d P_{n}(t)}{d t} 	&= \frac{d \Pi(\xi,t)}{d t} 
			= \frac{\partial \Pi}{\partial \xi}\frac{\partial \xi}{\partial t} + \frac{\partial \Pi}{\partial t} 						
			= \frac{\partial \Pi}{\partial t} - N^{\frac{1}{2}}\frac{d \mu}{d t}\frac{\partial \Pi}{\partial \xi}.
\end{align*}
Equating this to Equation~\ref{dPn} we obtain
\begin{align*}
\frac{\partial \Pi}{\partial t} - N^{\frac{1}{2}}\frac{\partial \mu}{\partial t}\frac{\partial \Pi}{\partial \xi}
		= 	&\left[N^{-\frac{1}{2}}\frac{\partial }{\partial \xi} +\frac{ N^{-1}}{2}\frac{\partial^2 }{\partial \xi^2} \ldots \right]\left(\alpha\left(N\mu + N^{\frac{1}{2}}\xi\right)\Pi\right)   \\
			&+\left[-N^{-\frac{1}{2}}\frac{\partial }{\partial \xi} + \frac{N^{-1}}{2}\frac{\partial^2 }{\partial \xi^2} \ldots \right]\left(f\left(w\left(\mu + N^{-\frac{1}{2}}\xi\right) +h\right)\left(N-\left(N\mu + N^{\frac{1}{2}}\xi\right)\right)\Pi\right) \\
		= 	&\left[N^{\frac{1}{2}}\frac{\partial }{\partial \xi} + \frac{1}{2}\frac{\partial^2 }{\partial \xi^2} \ldots \right]\left(\alpha\left(\mu + N^{-\frac{1}{2}}\xi\right)\Pi\right)\\
			&+\left[-N^{\frac{1}{2}}\frac{\partial }{\partial \xi} + \frac{1}{2}\frac{\partial^2 }{\partial \xi^2} \ldots \right]\left(f\left(w\left(\mu + N^{-\frac{1}{2}}\xi\right) +h\right)\left(1-\left(\mu + N^{-\frac{1}{2}}\xi\right)\right)\Pi	\right).
\end{align*}
For simplicity, we also now write $f\left(w\left(\mu + N^{-\frac{1}{2}}\xi\right) +h\right)$ as $f$.  Thus,
\begin{align*}
\frac{\partial \Pi}{\partial t} - N^{\frac{1}{2}}\frac{\partial \mu}{\partial t}\frac{\partial \Pi}{\partial \xi}
		=     &\left[N^{\frac{1}{2}}\frac{\partial }{\partial \xi} + \frac{1}{2}\frac{\partial^2 }{\partial \xi^2} \ldots \right]\left(\alpha\left(\mu + N^{-\frac{1}{2}}\xi\right)\Pi\right) \\
			&+\left[-N^{\frac{1}{2}}\frac{\partial }{\partial \xi} + \frac{1}{2}\frac{\partial^2 }{\partial \xi^2} \ldots \right]\left(f\left(1-\mu - N^{-\frac{1}{2}}\xi\right)\Pi\right)	\\
		=	&N^{\frac{1}{2}}\alpha\mu\frac{\partial\Pi}{\partial \xi} + \alpha\frac{\partial (\Pi\xi)}{\partial \xi} + \frac{\alpha\mu}{2}\frac{\partial^2\Pi}{\partial \xi^2} + \frac{\alpha}{2}N^{-\frac{1}{2}}\frac{\partial(\Pi\xi)}{\partial\xi} \\
			&+\left[-N^{\frac{1}{2}}\frac{\partial }{\partial \xi} + \frac{1}{2}\frac{\partial^2 }{\partial \xi^2} \ldots \right]\left(f\Pi\right) \\
			&-\left[-N^{\frac{1}{2}}\frac{\partial }{\partial \xi} + \frac{1}{2}\frac{\partial^2 }{\partial \xi^2} \ldots \right]\left(f\mu\Pi\right) \\
			&-\left[-N^{\frac{1}{2}}\frac{\partial }{\partial \xi} + \frac{1}{2}\frac{\partial^2 }{\partial \xi^2} \ldots \right]\left(fN^{-\frac{1}{2}}\xi\Pi\right)\\
		=	&N^{\frac{1}{2}}\alpha\mu\frac{\partial\Pi}{\partial \xi} + \alpha\frac{\partial (\Pi\xi)}{\partial \xi} + \frac{\alpha\mu}{2}\frac{\partial^2\Pi}{\partial \xi^2} + \frac{\alpha}{2}N^{-\frac{1}{2}}\frac{\partial(\Pi\xi)}{\partial\xi}  -N^{\frac{1}{2}}\frac{\partial(\Pi f)}{\partial \xi} 
+N^{\frac{1}{2}}\mu\frac{\partial (\Pi f)}{\partial \xi}\\ 
			&+ \frac{\partial (\Pi f\xi)}{\partial\xi} + \frac{1}{2}\frac{\partial^2(\Pi f)}{\partial \xi^2} - \frac{\mu}{2}\frac{\partial^2(\Pi f)}{\partial \xi^2} - \frac{N^{-\frac{1}{2}}}{2}\frac{\partial^2(\Pi f\xi)}{\partial \xi^2} + N^{-\frac{3}{2}}\ldots.
\end{align*}	
We note we can expand $f$ in terms of powers of $N^{-\frac{1}{2}}$ as follows
\begin{align*}
f\left(w\left(\mu+N^{-\frac{1}{2}}\xi\right)+h\right) 
			&= f\left(w\mu+h+wN^{-\frac{1}{2}}\xi\right)\\
			&= f(w\mu+h) + wN^{-\frac{1}{2}}\xi f^{'}(w\mu+h) + \frac{w^{2}\xi^{2}N^{-1}}{2!}f^{''}(w\mu+h)+\ldots.
\end{align*}
Writing $\hat{f}=f(w\mu+h)$ and $\hat{f}' = f'(w\mu+h)$ we find that
\begin{align*}
\frac{\partial(f\Pi)}{\partial\xi} 
				&=f\frac{\partial\Pi}{\partial\xi}+\Pi\frac{\partial f}{\partial \xi} \\
				&=\hat{f}\frac{\partial\Pi}{\partial\xi} + wN^{-\frac{1}{2}}\xi\hat{f}^{'}\frac{\partial\Pi}{\partial\xi}
				 + \Pi wN^{-\frac{1}{2}}\hat{f}^{'} + N^{-1}(\ldots) + \ldots\\
				 &=\hat{f}\frac{\partial\Pi}{\partial\xi} + N^{-\frac{1}{2}}\left(w\xi\hat{f}^{'}\frac{\partial\Pi}{\partial\xi}
				 + \Pi w\hat{f}^{'}\right) + N^{-1}(\ldots) + \ldots \\
\Rightarrow \frac{\partial^2(f\Pi)}{\partial\xi^2}
				&= \hat{f}\frac{\partial^2\Pi}{\partial\xi^2} + N^{-\frac{1}{2}}\left(w\xi\hat{f}^{'}\frac{\partial^2\Pi}{\partial\xi^2} 
				 + 2w\hat{f}^{'}\frac{\partial\Pi}{\partial\xi}\right)+ N^{-1}(\ldots) + \ldots.
\end{align*}
Similarly,
\begin{align*}
\frac{\partial(f\Pi\xi)}{\partial\xi}
				&= f\frac{\partial(\Pi\xi)}{\partial \xi} + (\Pi\xi)\frac{\partial f}{\partial\xi}\\
				&=\hat{f}\frac{\partial(\Pi\xi)}{\partial\xi} + wN^{-\frac{1}{2}}\xi\hat{f}^{'}\frac{\partial(\Pi\xi)}{\partial\xi}
				 +\Pi\xi wN^{-\frac{1}{2}}\hat{f}^{'} N^{-1}(\ldots) + \ldots \\
				&= \hat{f}\frac{\partial(\Pi\xi)}{\partial\xi} 
				 + N^{-\frac{1}{2}}\left(w\xi\hat{f}^{'}\frac{\partial(\Pi\xi)}{\partial\xi} +\Pi\xi w\hat{f}^{'}\right) N^{-1}(\ldots) + \ldots.
\end{align*}
Including these into our expansion we obtain
\begin{align*}
\frac{\partial \Pi}{\partial t} - N^{\frac{1}{2}}\frac{\partial \mu}{\partial t}\frac{\partial \Pi}{\partial \xi}		
		=	&N^{\frac{1}{2}}\alpha\mu\frac{\partial\Pi}{\partial \xi} + \alpha\frac{\partial (\Pi\xi)}{\partial \xi} +
		  \frac{\alpha\mu}{2}\frac{\partial^2\Pi}{\partial \xi^2}\\ 
		  &-N^{\frac{1}{2}}\left(\hat{f}\frac{\partial\Pi}{\partial\xi} + N^{-\frac{1}{2}}\left(w\xi\hat{f}'\frac{\partial\Pi}{\partial\xi} + \Pi w\hat{f}'\right)\right)\\ 
		  &+N^{\frac{1}{2}}\mu\left(\hat{f}\frac{\partial\Pi}{\partial\xi} + N^{-\frac{1}{2}}\left(w\xi\hat{f}'\frac{\partial\Pi}{\partial\xi} + \Pi w\hat{f}'\right)\right)\\
		  &+\hat{f}\frac{\partial\Pi\xi}{\partial\xi} + N^{-\frac{1}{2}}\left(w\xi\hat{f}'\frac{\partial(\Pi\xi)}{\partial\xi} + \Pi\xi w\hat{f}'\right)\\
		  &+\frac{1}{2}\left(\hat{f}\frac{\partial^2\Pi}{\partial\xi^2} + N^{-\frac{1}{2}}\left(w\xi\hat{f}'\frac{\partial^2\Pi}{\partial\xi^2}+2w\hat{f}'\frac{\partial\Pi}{\partial\xi}\right)\right)\\
		  &-\frac{\mu}{2}\left(\hat{f}\frac{\partial^2\Pi}{\partial\xi^2} + N^{-\frac{1}{2}}\left(w\xi\hat{f}'\frac{\partial^2\Pi}{\partial\xi^2}+2w\hat{f}'\frac{\partial\Pi}{\partial\xi}\right)\right)\\
		  &+N^{-\frac{1}{2}}\left(\ldots\right) \ldots \\
		=	&N^{\frac{1}{2}}\alpha\mu\frac{\partial\Pi}{\partial \xi} + \alpha\frac{\partial (\Pi\xi)}{\partial \xi} +
		  \frac{\alpha\mu}{2}\frac{\partial^2\Pi}{\partial \xi^2} - N^\frac{1}{2}\hat{f}\frac{\partial\Pi}{\partial\xi}\\
		  &-w\xi\hat{f}'\frac{\partial\Pi}{\partial\xi} - \Pi w\hat{f}' + N^\frac{1}{2}\mu\hat{f}\frac{\partial\Pi}{\partial\xi} + \mu w\xi\hat{f}'\frac{\partial\Pi}{\partial\xi}\\
		  &+\mu\Pi w\hat{f}' + \hat{f}\frac{\partial(\Pi\xi)}{\partial \xi} + \frac{\hat{f}}{2}\frac{\partial^2\Pi}{\partial\xi^2} - \frac{\mu}{2}\hat{f}\frac{\partial^2 \Pi}{\partial\xi^2}\\
			&+N^{-\frac{1}{2}}\left(\ldots\right)\ldots.
\end{align*}
We now group the terms in powers of $N^{\frac{1}{2}}$:
\begin{align*}
\frac{\partial \Pi}{\partial t} - N^{\frac{1}{2}}\frac{\partial \mu}{\partial t}\frac{\partial \Pi}{\partial \xi}
		=	&N^\frac{1}{2}\frac{\partial \Pi}{\partial \xi}\left(\alpha\mu - \hat{f} + \mu\hat{f}\right)
			+\frac{\partial(\Pi\xi)}{\partial\xi}\left(\alpha + \hat{f}\right)\\
			&+\frac{\partial^2 \Pi}{\partial\xi^2}\left(\frac{\alpha\mu}{2} + \frac{\hat{f}}{2} - \frac{\mu\hat{f}}{2}\right)\\
			&+\frac{\partial\Pi}{\partial\xi}\left(-w\xi\hat{f}'+ \mu w\xi\hat{f}'\right) +\mu\Pi w \hat{f}' -\Pi w\hat{f}' + N^{-\frac{1}{2}}(\ldots) 
			 + \ldots\\
		=	&N^\frac{1}{2}\frac{\partial \Pi}{\partial \xi}\left(\alpha\mu - \hat{f}(1-\mu)\right) 
			+\frac{\partial(\Pi\xi)}{\partial\xi}\left(\alpha + \hat{f}\right)\\
			&+\frac{\partial^2 \Pi}{\partial\xi^2}\left(\frac{\alpha\mu}{2} + \frac{\hat{f}}{2} - \frac{\mu\hat{f}}{2}\right)
			-\frac{\partial(\Pi\xi)}{\partial\xi}\left(w\hat{f}'(1-\mu)\right)
			+N^{-\frac{1}{2}}(\ldots).
\end{align*}
Choosing $\mu$ such that the macroscopic equation of the described process is satisfied (i.e.\ $\frac{d\mu}{d t} = -\alpha\mu + (1-\mu)\hat{f}$), we have the terms of order $N^{\frac{1}{2}}$ cancel out and we are left with an expansion in terms of $N^{-\frac{1}{2}}$. The leading order is then $N^{0}$ (but relative to the macroscopic solution $\mu$ this is $O(N^{-\frac{1}{2}})$.  Higher orders are now neglected as relative to the macroscopic $\mu$ they are order $N^{-1}$ (order of a single neurone). We are then left with the following linear Fokker-Planck equation
\begin{align*}
\frac{\partial \Pi}{\partial t} = \frac{\partial (\Pi\xi)}{\partial\xi}\left(\alpha + \hat{f} - w\hat{f}'(1-\mu)\right) + \frac{1}{2}\frac{\partial^2\Pi}{\partial \xi^2}\left(\alpha\mu + (1-\mu)\hat{f}\right).
\end{align*}
It was shown~\cite{vankampen} that the solution to this equation is gaussian so what we need to determine are the first and second moments.  For ease we will rewrite the above equation as
\begin{align*}
\frac{\partial \Pi}{\partial t} & = A(t)\frac{\partial (\Pi\xi)}{\partial\xi} + \frac{B(t)}{2}\frac{\partial^2\Pi}{\partial \xi^2}.
\end{align*}
The derivative of the first moment with respect to time is then given by
\begin{align}
\frac{\partial \left<\xi\right>}{\partial t} = \int_{\mathbb{R}}{\xi\frac{\partial \Pi}{\partial t}}d\xi 
&= A\int_{\mathbb{R}}{\xi\frac{\partial (\Pi\xi)}{\partial \xi}d\xi} + \frac{B}{2}\int_{\mathbb{R}}{\xi\frac{\partial^2 \Pi}{\partial\xi^2}d\xi}\nonumber
\\ &= A\left[\left.\xi^{2}\Pi\right|_{-\infty}^{\infty} - \int_{\mathbb{R}}{\xi\Pi d\xi}\right]  +\frac{B}{2}\left[\left.\xi\frac{\partial\Pi}{\partial\xi}\right|_{\infty}^{\infty} 
- \int_{\mathbb{R}}{\frac{\partial\Pi}{\partial\xi}d\xi}\right]\nonumber \\
&= -A\left<\xi\right> -\left[\frac{B\Pi}{2}\right]_{\infty}^{\infty} \nonumber\\
&= -A\left<\xi\right>. 	
\label{dXi}
\end{align}
Similarly, the derivative of the second moment with respect to time is given by
\begin{align}
\frac{\partial \left<\xi^2\right>}{\partial t} = \int_{\mathbb{R}}{\xi^{2}\frac{\partial \Pi}{\partial t}}d\xi
&= A\int_{\mathbb{R}}{\xi^2\frac{\partial (\Pi\xi)}{\partial\xi}d\xi} + \frac{B}{2}\int_{\mathbb{R}}{\xi^2\frac{\partial^2 \Pi}{\partial\xi^2}d\xi}\nonumber
\\ &= A\left[\left.\xi^{3}\Pi\right|_{\infty}^{\infty} - 2\int_{\mathbb{R}}{\Pi\xi^{2}}d\xi\right]
+ \frac{B}{2}\left[\left.\xi^2\frac{\partial\Pi}{\partial\xi}\right|_{\infty}^{\infty} 
- 2\int_{\mathbb{R}}{\xi\frac{\partial\Pi}{\partial\xi}}d\xi\right] \nonumber \\
&= -2A\left<\xi^{2}\right> - B\left[\left.\xi\Pi\right|_{\infty}^{\infty} - \int_{\mathbb{R}}{\Pi} d\xi\right] \nonumber \\
&= -2A\left<\xi^2\right> + B.
\end{align}
For both of these integrals many terms have vanished due to $\Pi(\xi,t)$ being a gaussian distribution and, in the limit, both the distribution and its derivative tend to zero quicker than any other factors involved.  We can similarly calculate the variance of the fluctuations as follows
\begin{align}
\sigma^2 &= \left< \xi^2 \right> - \left< \xi \right>^2 \nonumber \\
\Rightarrow \frac{\partial \left<\sigma^2\right>}{\partial t} &=
\frac{\partial \left<\xi^2\right>}{\partial t} - 2\left<\xi\right>\frac{\partial \left<\xi\right>}{\partial t}\nonumber \\
&= -2A\left<\xi^2\right> + B - 2\left<\xi\right>(-A\left<\xi\right>) \nonumber \\
&= -2A\left(\left<\xi^2\right> - \left<\xi\right>^2\right) + B \nonumber \\
&= -2A\sigma^2 + B.
\label{dsigma2}
\end{align} 
Using Equations~\ref{dXi} and \ref{dsigma2} with the correct values of $A$ and $B$ substituted in we can now compare the solutions of the following three equations with simulation results for the proposed model.  We note that as long as we know the initial distribution of active 
neurones then for the initial conditions we have $\left<\xi\right>_0 = \left<\xi^2\right>_0 = \left<\sigma^2\right>_0 = 0$ and $\left<\xi\right>$ will then
remain zero for all time.
\begin{align}
\frac{\partial\mu}{\partial t} &= -\alpha\mu + (1-\mu)\hat{f},\\
\frac{\partial \left<\xi\right>}{\partial t} &= -\left(\alpha + \hat{f} - w\hat{f}'(1-\mu)\right)\left<\xi\right>, \\
\frac{\partial \left<\sigma^2\right>}{\partial t} &= -2\left(\alpha + \hat{f} - w\hat{f}'(1-\mu)\right)\left<\sigma^2\right> +\left(\alpha\mu + (1-\mu)\hat{f}\right).
\end{align}
These in turn give the following equations for the mean, $A$, and variance, $A_\sigma$, of the number of active neurones:
\begin{align}
A &= N\mu + N^{-\frac{1}{2}}\left<\xi\right> = N\mu \quad \text{(assuming we know the initial number of active neurones)}, \\
A_{\sigma} &= N\left<\sigma^2\right>.
\end{align}
\bigskip

    
\section*{Acknowledgements}
  \ifthenelse{\boolean{publ}}{\small}{}
Timothy Taylor is funded by a PGR studentship from MRC, and the Departments of Informatics and Mathematics at University of Sussex. Caroline Hartley is funded through CoMPLEX (Centre for Mathematics and Physics in the Life Sciences and Experimental Biology), 
University College London. Istvan Z. Kiss acknowledges support from EPSRC (EP/H001085/1). P\'eter L. Simon acknowledges support from OTKA (grant no. 81403) and from the European Union and the European Social Fund (financial support to the project under the 
grant agreement no. T{\'A}MOP-4.2.1/B-09/1/KMR).

\theendnotes
 

\newpage
{\ifthenelse{\boolean{publ}}{\footnotesize}{\small}
 \bibliographystyle{bmc_article}  
  \bibliography{taylor_v1} }     


\ifthenelse{\boolean{publ}}{\end{multicols}}{}

\end{bmcformat}
\end{document}